\documentclass[a4paper,12pt]{article}
\usepackage{amssymb,amsmath,mathbbol,mathrsfs}
\usepackage[usenames,dvipsnames]{color}
\usepackage{hyperref}
\usepackage{xcolor}
\usepackage{stmaryrd}
\usepackage{authblk}
\usepackage{framed}
\usepackage{empheq} 
\usepackage{slashed}
\usepackage{hyphenat}
\usepackage{cite}
\usepackage{apacite}
\usepackage{natbib}
\usepackage{dsfont}
\usepackage{chngcntr}
\usepackage{enumerate}  
\usepackage{tikz-cd}

\usepackage{marginnote}    

\usepackage[left=.8in,right=.8in,top=.8in,bottom=.8in]{geometry}                  

\usepackage{sectsty} 
\allsectionsfont{\sffamily\mdseries\upshape} 
\usepackage{tocloft}

\makeatletter
\renewenvironment{abstract}{%
    \if@twocolumn
      \section*{\abstractname}%
    \else 
      \begin{center}%
        {\bfseries\sffamily\abstractname\vspace{\z@}}
      \end{center}%
      \quotation
    \fi}
    {\if@twocolumn\else\endquotation\fi}
\makeatother

\numberwithin{equation}{section}
\hypersetup{
	colorlinks=true,         
	linkcolor=MidnightBlue,          
	citecolor=BrickRed,        
	urlcolor=MidnightBlue            
}

\newcommand{\be}{\begin{equation}}
\newcommand{\ee}{\end{equation}}

\newcommand{\F}{{{\Phi}}}
\renewcommand{\d}{{\mathrm{d}}}
\newcommand{\D}{{\mathrm{D}}}

\newcommand{\G}{{\mathcal{G}}}

\newcommand{\pp}{{\partial}}

\newcommand{\Diff}{{\mathrm{Diff}}}

\renewcommand{\bar}{\overline}

\renewcommand{\hat}{\widehat}

\newcommand{\RR}{\mathds{R}} 

\newtheorem{defi}{Definition}

\newcommand{\cint}{{\int\kern-.87em{<}}}
\newcommand{\sint}{{\int\kern-.75em{\sim}}}
\newcommand{\fint}{{\int\kern-1.00em{\int}}}

\newcommand{\bb}{\mathbb}

\let\oldmarginpar\marginpar
\renewcommand\marginpar[1]{\oldmarginpar{\color{red}\raggedright\footnotesize #1}}

\title{  {\huge Same-diff?}\\  Conceptual similarities between gauge transformations and diffeomorphisms\\
{\large Part I: Symmetries and isomorphisms}
}
\author{Henrique de A. Gomes\footnote{\href{mailto:gomes.ha@gmail.com}{gomes.ha@gmail.com}} \\\it University of Cambridge\\ \it Trinity College, Cambridge, CB2 1TQ, UK\\
and\\
\it University College London\\
\it  Gower St, London WC1E 6BT
}

\begin{document}

\maketitle
\begin{abstract}
The following questions are germane to our understanding of gauge-(in)variant quantities and  physical possibility: how are gauge transformations and spacetime diffeomorphisms understood as symmetries, in which ways are they similar, and in which are they different? To what extent are we justified in endorsing different attitudes---nowadays called sophistication, haecceitism, and eliminativism---towards each?   This is the first of four papers taking up this question. 

 This first paper will discuss notions of symmetries and isomorphisms that will be used in the remaining papers in the series. There are several such notions in the literature and the question of how they mesh with empirical discernibility is a delicate one; even  the orthodox view that symmetries are empirically unobservable (even in principle) has  recently been challenged by \cite{Belot_sym}. Focusing on local field theories,  I will provide a precise definition of  dynamical symmetries in terms of the space of states of the theory at hand.  I will then apply the definition to Yang-Mills theories and general relativity and show that these symmetries correspond to automorphisms of `natural'  geometric structures: the small diffeomorphisms of the spacetime manifold and the small fiber-preserving diffeomorphisms of a fibered space. Finally, I will show that these automorphisms can be given a passive gloss, since they correspond 1-1 to the coordinate transformations of the underlying manifolds.
 
 
 
\end{abstract}

{\center \begin{quote}Same-diff [noun]: \textit{
an oxymoron, used to describe something as being the same as something else. Often used as an excuse for being wrong.} (Urban dictionary). 

\end{quote}}
\tableofcontents
\bigskip

\newpage

\section{Introduction}\label{sec:intro}

\subsection{Motivation}\label{sec:mot}
Gauge theories lie at the heart of modern physics: in particular, they constitute the standard model of particle physics.
Philosophers of physics generally accept as the leading idea of a gauge theory---or as the main connotation of the phrase `gauge theory’---that it involves a formalism that uses more variables than there are physical degrees of freedom in the system described; and thereby more variables that one strictly speaking needs to use. Hence the common soubriquets: `descriptive redundancy’, `surplus structure’, and more controversially, `descriptive fluff’ (e.g. \cite{Earman_gmatters, Earman2004}). 

Although the main idea and connotation of descriptive redundancy has been endorsed by countless presentations in the physics literature, some celebrated philosophers, such as \citet{Healey_book} and \citet{Earman_gmatters} among others, have gone beyond this connotation, and defended a stronger,  \emph{eliminativist view}. The view is that  gauge symmetry must be  `eliminated' before determining which models of a theory represent distinct physical possibilities, on pain of radical indeterminism.\footnote{We will more fully describe what is expected of eliminativism, and its alternatives, in \cite{Samediff_1a}. For now, I take a gauge symmetry to be `eliminated' in the sense required if there exists a second theory, with different ontological commitments, which is empirically equivalent to the first, and which has no local gauge symmetry.} 
For them,  the connotation of `fluff' is that it can have no purpose. 

But radical indeterminism also threatens theories such as general relativity, embodying diffeomorphism symmetry. Does this symmetry arise from `descriptive redundancy' in the same way as, it is claimed, gauge transformations do? Should we construe the inference from models to reality similarly in the two theories?  In this paper, I will show that, under a specific definition of dynamical symmetries, those of both general relativity and Yang-Mills theory  can be understood to arise from descriptive redundancy. But here I will not attempt to elevate this conclusion to a criterion, regimenting when symmetries can be understood in this way, as descriptive redundancies. That will be the job of the second \citep{Samediff_1a} and third \cite{Samediff_1b} papers in the series.

In this first of four papers analysing the similarities and distinctions between the gauge symmetries of Yang-Mills theory and the spacetime diffeomorphisms of general relativity, I will set up the formal background, the basic physical interpretation, and the basic definitions to be used in the remaining three. 
The second and third paper, \cite{Samediff_1a, Samediff_1b} will analyse more formal aspects of the comparison between the gauge symmetries of Yang-Mills theory and the spacetime diffeomorphisms of general relativity.  They will also give general  desiderata for other theories  to admit a perspicuous interpretation, in a similar way as  Yang-Mills and general relativity do. \cite{Samediff_1a} focuses on topics that are more metaphysical and concern the philosopher more than the physicist, while \cite{Samediff_1b} focuses on conceptual matters that  are nearer to the heart of physicists. The fourth paper in the series, \cite{Samediff_2}, will analyse more detailed aspects  of this comparison, such as the degree of non-locality of the two theories.

This paper sets the standard for the following ones, by construing the different types of interaction---e.g. electromagnetic---geometrically, as on a par with how general relativity describes gravity. I describe how both the fundamental fields of these theories  encode structural, or relational,  properties, that arise from comparisons. If spacetime geometry is about the external distance between spacetime points, the principal bundle geometry is about the internal `distances' (or rather, rotations) between the charges of particles. And in even fewer words: general relativity is about the external geometry, whereas Yang-Mills theory is about the internal geometry. 

With that construal, I hope to erase, or at least weaken, any prejudice the reader may harbor about fundamental conceptual differences between the symmetries of  general relativity and Yang-Mills theory. 

\subsection{Roadmap}
 Here is a brief outline about how we plan to proceed. In Section \ref{sec:syms} I   will provide a detailed definition of symmetries, including infinitesimal symmetries. When we apply the general definition of symmetries to the functions that are responsible for endowing the theory with dynamical content---i.e. a Hamiltonian or an action functional---we arrive at the \emph{empirical unobservability thesis}: that symmetry-related models are empirically indiscernible.\footnote{This thesis will be defended at a more technical level in the third paper in the series, \cite{Samediff_1b}, once we have developed the necessary tools.}   Interpreting these symmetries as the isomorphisms of some category will enable me to give a rough outline of the doctrines of \emph{eliminativism and sophistication}, which will be main topics in the following paper in the series, \cite{Samediff_1a}.  In Section \ref{sec:attitudes} we provide a brief introduction to the mathematical formalism of the theory of general relativity.  
 
 In Section \ref{sec:PFB} we do the same for gauge theory, but with greater attention to detail, since the theory is less familiar to the philosopher of physics. As to  the dynamical symmetries of general relativity and Yang-Mills, I will  display an exhaustive set of symmetries that are infinitesimal, or connected to the identity, according to the  definition of Section \ref{sec:syms},  in each theory. I will do this in Sections \ref{sec:attitudes} and  \ref{sec:PFB}, respectively.\footnote{Most of the literature on the topic does \emph{not} show this: they merely present some set of transformations that are symmetries under a given definition---cf. \cite{Torre_GRsym} for an exception.}  In Section \ref{sec:PFB}, we  encounter two types of symmetries: ones that can be interpreted via the isomorphisms of some natural geometric structure, and ones that are just a postulated mathematical transformation. In parallel, philosophers of physics are accustomed to the active interpretation of symmetries---as isomorphisms of some natural geometric structure--- whereas more pragmatic physicists tend to construe symmetries passively, as mere postulated changes between coordinate systems. 
 
 In Section \ref{sec:p-a}, I defuse this tension, by providing a one-to-one correspondence between the (infinitesimal) symmetries defined in Section \ref{sec:syms} and passive changes of coordinates of the natural geometric structures underlying general relativity and Yang-Mills theory. This resolution allows us to see the dynamical symmetries of both theories as descriptive redundancies.  
 
 Finally, I note that the basic Yang-Mills field that lends itself to the geometric interepretation is \emph{not} a field on spacetime; it is a field on some other (fibered) manifold and requires  coordinate charts for representations on spacetime.  Therefore, to finish the side-by-side comparison of Yang-Mills and general relativity, we would like to describe the  Yang-Mills fields as on a par with the abstract metric tensor field, as fields on spacetime and without the use of coordinate charts.  We provide this interpretation by construing the basic fields of Yang-Mills theory as sections of the bundle of connections (or Atiyah-Lie bundle). Since this construction is overly technical, we leave it to Appendix  \ref{sec:Atiyah}.\footnote{This construction will be important in the fourth paper in the series, \cite{Samediff_2}.}  In Section \ref{sec:conclusions_soph} we conclude.

\section{Dynamical symmetries}\label{sec:syms}
To begin our more formal investigation,  I must  provide a  formal definition of symmetries. This may seem like a straightforward task, but it is far from it. The intuitions we commonly have about symmetries clash with most attempts of formalization (as discussed by \cite{Belot_sym}). So we tread carefully, and define symmetries more flexibly than is usually done. 
This brief treatment already allows us to ask interesting questions, about the interpretation of symmetries, and about symmetry-related models. 

In its broadest terms, a symmetry is a transformation of a system which preserves the values of a relevant (usually large) set of physical quantities. Of course, this broad idea is made precise in various different ways: for example as a map on the space of states, or on the set of quantities; and as a map that must respect the system's dynamics, e.g.  by mapping solutions to solutions or even by preserving the value of the Lagrangian functional on the states.

 In Section \ref{sec:syms_tech} I will  provide the definitions about symmetries that we will be using throughout this paper. In Section \ref{sec:emp_unob} I will argue that, applying this notion to the generators of dynamical evolution, it is plausible to infer that symmetry-related models are empirically indiscernible. Section \ref{sec:syms_isos} discusses the relation between the idea of symmetries explored in the previous Section and the existence of an appropriate mathematical structure that encodes those symmetries as its isomorphisms. 
Given the tools of Section \ref{sec:syms_isos}, Section \ref{sec:soph_elim} briefly discusses the  doctrine of \emph{structuralism} and its relation to the  reductive understanding of symmetry-related models, called \emph{eliminativism}, that will be an important thread in the following papers in the series.

\subsection{Technical considerations about symmetries}\label{sec:syms_tech}
The intuitive idea of  dynamical symmetries is that they are transformations acting on the models, or solutions, of a given theory, such that the  models that they relate are empirically indiscernible according to that theory. The intuition is helpful, but  nailing down symmetry more precisely  is a  challenge. For instance: defining a dynamical symmetry as any transformation that takes each solution of the equations of motion of a theory to another solution is far too weak: it would imply that any solution is related by a symmetry to any other. And there are other problems. For instance:  models which we would intuitively take to depict physically distinct situations may  nonetheless be symmetry-related, depending on the notion of symmetry; and it is also false that  empirically identical situations are always symmetry-related according to every account of symmetry.\footnote{Nor is it straightforward to nail down what ``preserving the form'' of an equation really means. But this can be achieved by using the formalism of jet bundles: see, for example \cite{Weatherall_jet}.}  

Examples illustrating the above problems---and more---are described in \citep{Belot_sym}, which expounds  the obstacles towards a general definition. Different authors have risen to Belot's challenge, of providing a general account of symmetry that is   coherent and yet non-circular. For instance,  \cite{Wallace2019} requires symmetries to be realizable as transformations of subsystems of the universe, while \cite{Fletcher2018} requires other non-physical, epistemic criteria. I want to avoid the discussion of subsystems and would prefer an explanation based on mathematical/physical criteria.  
 So for now, I give what I believe to be a plausible definition of symmetries, that disallows some but probably not all of \cite{Belot_sym}'s counter-examples.

   Let $\F$ be the space of models of the theory. Models are supposed to be complete descriptions of the world, according to the given theory. Here the word `world' is deliberately ambiguous: it can refer to an  instantaneous state or to an entire history.\footnote{This first definition excludes subsystems. I discuss these in \cite{DES_gf}, and, in more generality, in \cite{Gomes_bdary}.}   
 And `instantaneous state' is also ambiguous: one may understand an instantaneous description to include or not include information about rates of change---theories whose models are  states in phase space include this information and those whose models are complete instantaneous configurations do not.  Models of instantaneous states of affairs (in both senses) will here be dubbed \emph{states} of the \emph{universe}; and I will keep using  `world' and `model' as the more inclusive terms: both can apply  to descriptions of entire histories or of instantaneous states.

Now, each physical theory will postulate some mathematical structure for its models. 
For example, in non-relativistic mechanics, we could have \emph{each model} be a configuration of $N$ point particles in Euclidean space, $\RR^3$. So each model is endowed with both the differentiable and vector space structure of $\RR^3$, which can be used in formal manipulations. Now this mathematical structure \emph{of each model} is reflected in a different level of mathematical structure \emph{for the entire space of models}, $\F$. In this example, the space of models---taken as instantaneous states without information about rates of change---is  configuration space, which is isomorphic to $\RR^{3N}$. So, while the linear and smooth structure of $\RR^3$ belongs to each model, and we use it for important operations such as taking derivatives, we also require the smooth structure of configuration space in order to do variational calculus, or to give a Hamiltonian formulation of the theory. Or similarly, the symplectic structure of a given theory can be seen as a structure on the state space $\F$; this structure does not inhere in each model (which can itself have a lot of structure, in particular  in the case of field theories: e.g. for a model of a history in general relativity, i.e. a model of spacetime, the structure  of a semi-Riemannian manifold). 

 In field theories, the space of models ${\F}$ is usually endowed with an (infinite-dimensional) topological structure that allows definitions of neighborhoods of models,  differentiable one-parameter families of models, etc. And as discussed in the previous paragraph, we will usually endow it with further structure: smooth, symplectic, etc.\footnote{An important question here is: in what sense does the mathematical structure of the models constrain or determine the mathematical structure of $\F$? For example, in \cite[Ch. 10]{Ringstrom_book}, it is argued that other criteria, such as stability of solutions of the theory, have the power to largely determine the appropriate topology of $\F$. But I do not aim to answer this complicated question in general. } 
Of course, using these further, e.g. topological, structures, $\F$ becomes an infinite-dimensional manifold. But I would like to reassure the concerned reader on this point: infinite and finite dimensional geometries differ in various details, but much of the abstract geometrical reasoning that we are familiar with in the finite case extends to the infinite one.\footnote{\cite{Michor} have a general approach to geometry that is based on curves and their differentiability as embedded in arbitrary spaces; and for many of the geometrical objects and intuitions of the finite-dimensional case,  the approach builds  bridges towards the infinite-dimensional. Another useful source, that develops differential geometry in the infinite-dimensional case by replacing $\RR^n$ as the image of local charts of manifolds  by more general Hilbert or Banach vector spaces, is \citep{Lang_book}. One useful rule of thumb about generalizing mathematical theorems is the following: theorems of finite-dimensional geometry whose proof requires some sort of integration are not straightforwardly extendible, whereas those that do not require integration are relatively easily extendible.\label{ftnt:inf_dim}}

Thus, in sum,  each of the models and also ${\F}$ are endowed with mathematical structure. Now we can define a general notion of symmetry.

\begin{defi}[$S$-symmetry]\label{def:sym} Let $S$ be some quantity on the system, represented as a real function on ${\F}$ that respects these structures (e.g. is   smooth, linear, etc.).  Then a transformation $\Theta:{\F}\rightarrow {\F}$ is an $S$-symmetry iff $\Theta$: \begin{enumerate}[(a)]
 \item respects the mathematical structure posited for ${\F}$ (e.g.  smooth, linear, symplectic, etc.);  
\item is definable without fixed parameters from ${\F}$, i.e. all models enter as free variables in the  transformation $\Theta$; and
 \item $\Theta$ preserves the values of $S$: for any model $m\in \F$, $S(\Theta(m))=S(m)$. \end{enumerate}
\end{defi}
 
  Note  that a transformation $\Theta$ that only preserves the value of $S$ at a subset of models is not an  $S$-symmetry. A symmetry transformation respects   the mathematical structure posited for ${\F}$ \emph{and} preserves the value of a function $S$  on ${\F}$. So, for example, given some such structure,  e.g. a symplectic form $\Omega$ (in which case ${\F}$ is a smooth manifold, infinite-dimensional in the case of field theories and finite-dimensional for  particle mechanics), and a Hamiltonian $H$ that is a real-valued function on ${\F}$,  then then (using the asterisk, as usual, for drag-along and pull-back), item (a) implies $\Theta^*\Omega=\Omega$, and item (c) implies $\Theta^*H:=H\circ \Theta=H$. 
  
  The purpose of item (b) is to disallow `spurious' symmetries. That is, in the same way that we would not like any two solutions of the equations of motion to be related by a symmetry, we do not want to say that all of the states with the same $S$ are related by $S$-symmetries. Item (b) disallows such gerrymandered $\Theta$'s.\footnote{However, I should note that, in most cases, respecting the smooth structure of $\F$ as per item (a) would already disallow crudely  gerrymandered situations; but one can certainly create examples that satisfy (a) and would be disallowed by (b). } \\
  
 Given Definition \ref{def:sym} it is, for precisely formulated theories, usually straightforward to exhibit some symmetries. But  the challenge is not to determine that some set of transformations---say the diffeomorphisms---are symmetries of some theory, say general relativity. The challenge is to find \emph{all} the symmetries, under some given definition and formulation of the theory. In that respect, it is easier to work with infinitesimal symmetries, to which we now turn.  

\paragraph{Infinitesimal symmetries}
  In this paper I will  be mostly interested in local symmetries of field theories, and in symmetries that could arise in either the Hamiltonian or the Lagrangian framework. That is, I am interested in symmetries that are  continuous and connected to the identity transformation, and in the case where the space of models ${\F}$ has at least a topological structure. 
  
   Specializing to those cases,  let $S$ be some quantity on ${\F}$, as above.  
   \begin{defi}[Infinitesimal $S$-symmetry]\label{def:inf_sym}   A vector field $\mathcal{X}$ on ${\F}$  generates an infinitesimal $S$-symmetry, iff:
   \begin{enumerate}[(a)]
 \item $\mathcal{X}$ respects the structure of ${\F}$ (e.g. its flow is continuous, smooth, symplectic,  etc.); 
 \item $\mathcal{X}$ is definable without fixed parameters from ${\F}$, i.e. all models enter as free variables in the  argument of $\mathcal{X}$; and
 \item  $\mathcal{X}$ preserves the values of $S$: for any model $m$, $\mathcal{X}[S](m)=0$.\footnote{Here, for any 1-parameter curve of models $m(t)$ such that $\frac{d}{dt}_{|t=0}m(t)={\cal X}_m$ and $m(0)=m$, this is taken as $\mathcal{X}[S](m):=\frac{d}{dt}_{|t=0}(S(m(t))$.}
 \end{enumerate}
\end{defi}

  When an infinitesimal $S$-symmetry can be integrated for parameter time $t$, we have finite symmetries generated by the flow of $\cal X$:  $\Theta_{\mathcal{X}}^t:{\F}\rightarrow {\F}$, such that  (omitting $\cal X$ and $t$): $S(\Theta(m))=S(m)$.  
 
 Infinitesimal symmetries are generically much more tractable than the full group of symmetries; and, even in field theory, given $S$, they can often be found algorithmically, e.g. as kernels of certain integro-differential operators: which is how we will determine them from the Einstein-Hilbert and the Yang-Mills action functionals, in Sections \ref{sec:attitudes} and \ref{sec:PFB}, respectively.

These definitions downplay the role of the dynamical equations of motion of a given theory. But we can include the dynamical content of symmetries by equating $S$ with an \emph{action functional}.  Such an action functional provides a more complete characterization of the dynamics of a given theory than do the equations of motion, since it can be used as a starting-point for quantization within either the Lagrangian or Hamiltonian formalisms; (and it also yields the classical equations of motion in a straightforward manner). 
So, for almost the entirety of this paper, the quantity $S$ for the $S$-symmetries will be identified with the action functional. And so here $m\in \F$ is a history, for which I will suppress boundary conditions in the elementary notation, and I will write $\varphi$ for $m$, to match field theory notation---which will be my focus.


Thus  we endow $\F$ with a (infinite-dimensional) manifold-like structure of its own; and take dynamics to be obtained from a variational principle. That is, given an action functional on this space: $S:\F\rightarrow \RR$, the extremization requirement $S[\varphi+\delta \varphi]-S[\varphi]=0$ for all directions (or vector fields) $\delta \varphi\in T_{\varphi}\F$, gives rise to the equations of motion, as conditions on the `base point' $\varphi$. Moreover, certain vector fields on $\F$ may leave $S$ invariant,  e.g.  $S[\varphi+\widehat{\delta \varphi}(\varphi)]-S[\varphi]=0$, for all $\varphi$, where $\widehat{\delta \varphi}:\F\rightarrow T_\varphi\F$ is a smooth vector field on this infinite-dimensional field space, $\F$, that, importantly, obeys supposition (b) from Definition \ref{def:inf_sym}. 

 So much by the way of summarizing the situation in classical physics. Turning to quantum mechanics:   barring the existence of anomalies (which arise from the lack of invariance of the path integral measure and/or regularization), symmetries of an action are straightforwardly translated into quantum symmetries. Thus it is natural to take this as a more fundamental notion  than symmetry of the equations of motion. This is the first reason to prefer the notion of infinitesimal symmetries; there are two more, as I now explain.  In the path integral formalism,  infinitesimal symmetries constitute degenerate directions for the propagator. This implies that two nearby states lying along such a degenerate direction should count as physically the same. In order that the contribution of states in these directions are not counted independently towards a given transition amplitude, we are motivated to identify  them as being physically identical. Thirdly,  infinitesimal symmetries are the only local symmetries that can arise in the Hamiltonian formalism (as discussed in the third paper, \cite{Samediff_2}). Thus we take Definition \ref{def:inf_sym} as  more fundamental. 
   
 \subsection{Empirical unobservability}\label{sec:emp_unob}

An $S$-symmetry relates empirically indistinguishable models if $S$ captures all the empirically accessible quantities.\footnote{Here I do \emph{not} use `empirical' to denote the traditional positivist and post-positivist `meter-readings' or `no-special-training-neeed for the judgment', or `the sheer look'---a very common denotation in the  literature about the theory-observation distinction of the past fifty years. I use it to denote `in-principle-observable', in a very encompassing sense of `in-principle'.\label{ftnt:empirical} } Theories are their own arbiters of empirical (in)discernibility (cf. \citep{ReadMoller}),\footnote{Einstein made this very point to Heisenberg. Here is how  \cite[p.63]{Heisenberg_dial} described the interaction: 
\begin{quote} I said ``We cannot observe electron orbits inside the atom...Now, since a good theory must be based on directly observable magnitudes, I thought it more fitting to restrict myself to these, treating them, as it were, as representatives of the electron orbits." But Einstein protested: ``But you don't seriously believe that none but observable magnitudes must go into a physical theory?". In some surprise, I asked ``Isn't that precisely what you have done with relativity?". 
``Possibly I did use this kind of reasoning," Einstein admitted, ``but it is nonsense all the same....In reality the very opposite happens. It is the theory which decides what we can observe." \end{quote}
The topic prompts one to consider the Kantian view that there are certain `a priori' elements of any given theory, which must be assumed if the theory is to be empirically significant; for this reason these a priori elements cannot be cannot be empirically tested in the same ways as the other elements in the theory. Along roughly the same lines, the relativized, or dynamic a priori, as elaborated in \citep{Friedman_dyn}, provides an updated version of the position, that can, Friedman argues, withstand the weight of evidence from the history of modern science against its Kantian forebear.} so different theories may have different  $S$'s being sufficient for empirical indiscernibility. But for  all theories of modern physics,  taking $S$ as the Hamiltonian or the action functional will be enough for our purposes.\footnote{Boundary conditions are here taken as features of ${\F}$, jointly with  the other mathematical structure delineated above. One of the most notable counter-examples of \cite{Belot_sym} is the Lenz-Runge symmetry, which preserves the equations of motion of a Newtonian two-body problem, but does not preserve features we take to be observable, such as the orbit eccentricity. We could disallow these symmetries by including eccentricity as one of our quantities $S$, but, in this case, this is not necessary. For Lenz-Runge symmetry is not an $S$-symmetry when $S$ is the action functional, since the action is not preserved by that symmetry: it is only preserved up to a boundary term that is non-vanishing. Galilean boosts are similarly excluded: they introduce a boundary term. A milder definition of $S$, which allows arbitrary boundary terms, is also a possibility, and indeed, in general relativity in the presence of boundaries, it is necessary to allow boundary terms  in order that  infinitesimal diffeomorphisms count as symmetries of the standard  Einstein-Hilbert action of the theory. } 

To be more precise: it is not that I believe that the action or Hamiltonian somehow encompasses all physical quantities for a given theory: it is rather that I endorse \emph{the unobservability thesis} of \cite{Wallace2019}. That is,  take the generator of the dynamics to be the Hamiltonian or the action functional. It is only the variation of these functions that dictate the evolution: e.g. through a Poisson bracket or a variational principle. Thus, if these quantities do not vary when the basic variables are acted on by a transformation satisfying (a) and (b) of Definition \ref{def:inf_sym},  there is a rigorous sense in which the very dynamics of the theory are preserved by the set of transformations. Now, we can further assume that empirical access to a physical system (see footnote \ref{ftnt:empirical}), in particular a physical process of observation,  is itself a dynamical notion. Thus a dynamical symmetry cannot have consequences for what is observable  when that symmetry encompasses
the physical processes involved in a  measurement. From these two suppositions, it is not far-fetched to conclude that quantities or properties that are symmetry-\emph{variant} cannot make a difference to a dynamical process. Or put differently:  the values of such quantities cannot be inferred from dynamical processes; and in particular, by certain types of observation.  That is,  under certain assumptions about the measurement process,
 the unobservability thesis states  that quantities or properties that have values that vary under  transformations of the
system that preserve all dynamical facts---e.g. the equations of motion or the quantum transition amplitudes---are unobservable, because they cannot be ascribed dynamical significance. We will have more to say about this in the third paper in the series, \cite{Samediff_1b}.
 
\subsection{Symmetries as isomorphisms}\label{sec:syms_isos}

 But, as will be discussed at length in this paper and its sequel: if we are to judge symmetry-related models as representing the same physical possibility, it makes sense to seek a type of physical and mathematical structure that clearly represents the quantities that are symmetry-invariant.\footnote{ It is this idea that motivates our first desideratum for \emph{sophistication}, in \cite{Samediff_1a}, i.e. (i): that symmetries be mathematically induced by the automorphisms of some natural geometric structure. We will briefly introduce the doctrine in Section \ref{sec:soph_elim}.}

 The first step in realizing this idea is naturally conveyed in the category theoretic framework (cf. footnote \ref{ftnt:category} below): we identify symmetries with the isomorphisms of some structure, as represented in a category in which the objects are the models of the theory.
 That is, since item (c) in Definition \ref{def:inf_sym} implies that symmetries can be composed,  
 we  demand that symmetries, acting on models,  form a groupoid, i.e. a category in which every arrow is an  isomorphism, with the objects of the category being the models, i.e. the elements of   ${\F}$.\footnote{The most important characteristic of category theory is its focus on isomorphisms and transformations between
mathematical objects that preserve (some of) their internal structure. For instance, these isomorphisms could be group homomorphisms in the category of groups, or linear maps in the category of
vector spaces. More precisely, given a
category $\mathcal{C}$, a morphism $f:A\rightarrow B$ is an isomorphism between objects $A$ and $B$ if and only if there is another morphism $f^{-1}:B\rightarrow A$ such that $f\circ f^{-1}=\mathsf{Id}_A$ and $f^{-1}\circ f=\mathsf{Id}_B$. And a property $P$ is structural, just in case $P(A)$ iff $P(f(A))$ for all isomorphisms $f$.  Another important type of mapping are the functors between different categories. This is, essentially, a mapping of objects to objects and arrows to arrows that preserves 
the categorical properties in question. Such functors  are crucial  for comparing
the objects of different mathematical categories. A groupoid is a category in which every arrow has an inverse in the above sense, i.e. every arrow is an isomorphism. An automorphism is an isomorphism that has the same object as its domain and target: $f:A\rightarrow A$. Thus, for instance, to take the example of Section \ref{sec:attitudes}, the automorphisms of the differentiable manifold $M$---the diffeomorphisms $\Diff(M)$---will induce, through pull-back, the isomorphisms of  Lorentzian metrics on $M$, the category of objects Lor$(M)$. The automophisms of a Lorentzian metric are those diffeomorphisms that preserve the metric, i.e. they are isometries $f^*g_{ab}=g_{ab}$ (and so, for generic objects of Lor$(M)$, they are just the identity). \label{ftnt:category} } 

Lest these definitions remain strictly mathematical,  I make explicit, albeit in a general way, their relation to the physical world (or better: that part of the physical world the theory aims to describe). We assume there is a set of possible worlds so that  our theory maps each orbit---under the symmetries---of models to a class of physically equivalent worlds, in a 1-1 manner (see \cite{Jacobs_rep} for more on this relation). So every world is described by some model, and two worlds are physically equivalent iff they are described by isomorphic models. 

 This initial construal of symmetries  is closest to what \cite[p. 3]{Wallace2019} dubs the `representational strategy', which ``builds the representational equivalence of symmetry-related models into the definition [of symmetry], usually by requiring that symmetries
are automorphisms of the appropriate mathematical space of models (hence
preserve all structure, and thus all representation-apt features, of a model)''. 

But note the flexibility of the formalism: we have not specified any independent definition of invariant structure; for now invariant structure is just what is common to the symmetry-related models. I will postpone to  \cite{Samediff_1a} whether this construal of symmetries as isomorphisms is apt for sophistication.\\

 All the symmetries investigated here and in the following papers,  obeying Definition \ref{def:inf_sym}, will be (when exponentiated) represented as groups (which could be infinite-dimensional), denoted $\G$, such that, given the space  of models of a theory, ${\F}$, there is an action  of $\G$ on ${\F}$, a map ${\F}:\G\times {\F}\rightarrow {\F}$, that preserves the action functional, and so that each $g\in \G$ gives an $S$-symmetry.
 More formally: there is  a structure-preserving map, $\mu$, on $\F$ that can be characterized  element-wise,  for $g\in \G$ and $\varphi\in \F$, as follows:
 \begin{eqnarray}\mu:\G\times \F&\rightarrow&\F\nonumber\\
(g,\varphi)&\mapsto&\mu(g, \varphi)=: \varphi^g. \label{eq:group_action}
\end{eqnarray}
Since each $g\in \G$ defines a symmetry, the action is such that, as per clause (c)  in the Definitions above, $S(\varphi^g)=S(\varphi)$, for all $\varphi$ and $g$. 

The symmetry group partitions the state space into equivalence classes by an equivalence relation, $\sim$, where $\varphi\sim \varphi'$ iff for some $g$, $\varphi'=\varphi^g$. We denote the equivalence classes under this relation by square brackets $[\varphi]$ and  the orbit of $\varphi$ under $\mathcal{G}$ by $\mathcal{O}_\varphi:=\{\varphi^g, g\in \mathcal{G}\}$. Note here that  though there is a one-to-one correspondence between $[\varphi]$ and $\mathcal{O}_\varphi$, the latter is  seen as an embedded manifold of $\F$, whereas the former exists abstractly, outside of $\F$ (see Figure 1).
\begin{figure}[h]
  \centering
  \includegraphics[width= 0.5\textwidth]{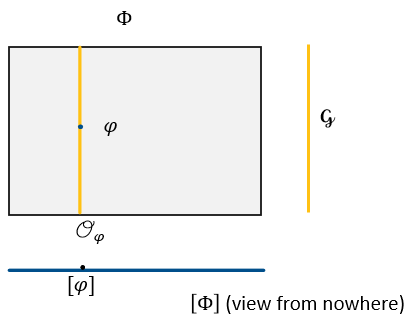}
  \caption{
The space of states, `foliated' by the action of some group $\mathcal{G}$ that preserves the value of some relevant quantity, $S$, and the space of equivalence classes. In field theory, when $S$ is the value of the action functional, each of these spaces---$\F$, $\mathcal{G}$, and $[\F]$---is an infinite-dimensional manifold. 
  }\label{fig:PFB_mine}
\end{figure} More mathematically: were we to write the canonical projection operator onto the equivalence classes, $\mathsf{pr}:{\F}\rightarrow {\F}/\G$, taking $\varphi\mapsto [\varphi]$, then  the orbit ${\cal O}_\varphi$ is the pre-image of this projection, i.e.  $\mathcal{O}_\varphi:=\mathsf{pr}^{-1}([\varphi])$. 



 Tacitly endorsing these extra assumptions about symmetries,  we call  $[\varphi]$ the  \textit{physical state}, and $\varphi'\in \mathcal{O}_\varphi$  its \textit{representative} (when there is no need to  emphasise that $\varphi$ involves a choice of representative, we call it just `the state'  for short). We call the collection of equivalence classes, $[{\F}]:=\{[\phi], \phi\in{\F}\}$,  \emph{the physical state space}. As written, this is an abstract space, i.e. defined implicitly by an equivalence relation, or as  certain classes of isomorphic models, under the appropriate notion of isomorphism.  It is, in a perspectival analogy, `a view from nowhere'.

 \subsection{Structuralism in physics, summarized}\label{sec:soph_elim}
There is an important distinction  between the objects represented by the models and the structure that is represented by the isomorphism classes of the models.\footnote{It is unfortunate that the label `stucturalism' has already been applied to the relational-substantivalist debate with a different meaning: for \cite{Ladyman_PII}, structuralism carried connotations of eliminativism. Here, it does not. \label{ftnt:Ladyman}}  
 

 In  physics, the distinction becomes more salient in the context of \emph{determinism}.  
In the case of theories with `time-dependent' symmetries---such as Yang-Mills theory and general relativity---determinism can only be secured for the equivalence classes, $[\varphi]\in [\F]$,  not for the states $\varphi\in \F$ (see e.g. \cite{Wallace_LagSym, Earman_det}). 

But, as in pure  mathematics, we usually cannot explicitly express the structure encoded by $[\varphi]$ (at least not without significant pragmatic burden or explanatory deficit); we can  do so only implicitly, by pointing to the isomorphism  classes, or by selecting representatives of those classes. Thus we enter debates about structuralism within physics. With the jargon introduced, we can briefly revisit some of the definitions glossed in Section \ref{sec:intro}:

 \emph{Eliminativism} about symmetries  seeks a new theory with an  \emph{intrinsic parametrization of $[\F]$} that makes no reference to the elements of $\F$. In other words, eliminativism seeks to render all and only  the  physically significant structure of the old theory  as the primary objects of a new theory, thus securing   physical determinism by jettisoning  representational redundancy.

\emph{Sophistication}, in broad terms,  rejects eliminativism while maintaining a commitment to structuralism as an abstract---often higher-order, in the logical sense of requiring quantification over properties and relations---characterization of the ontology, often under the label of `Leibniz  equivalence'. This position claims that an intrinsic parametrization of $[\F]$ is \emph{not} required for an ontological commitment only to members of $[\F]$ (see \cite{Dewar2017}); the broad idea is to use arbitrary members of $\mathcal{O}_\varphi$ as opposed to $[\varphi]$. We will have much more to say about this doctrine---and in defence of it!--- in the remaining papers in this series, once we have introduced the theories we want to apply the doctrine to.

\section{Diffeomorphisms in general relativity}\label{sec:attitudes}\label{sec:diff_sym}

This Section will be briefer than the following one, on gauge symmetry, since the intepretation of redundancy in general relativity is less controversial than  in gauge theory. 



I will take general relativity in the metric formalism, where the most general models of the theory, sometimes labeled \emph{kinematically possible models} (KPMs) (so as to avoid confusion with those models that satisfy the equations of motion, which are labeled \emph{dynamically possible models} (DPMs)), are given by the tuples: $\langle M, g_{ab}, \nabla, \psi\rangle$. Here $M$ is a smooth manifold,   $g_{ab}$ is a Lorentzian metric (a $(0,2)$-rank tensor with signature $(-,+,+,+)$);  $\nabla$ is a covariant derivative operator, and $\psi$ represents some distribution of matter and radiation. I will assume $\nabla$ is the the unique Levi-Civita one, i.e. obeying $\nabla_c g_{ab}=0$. I will call the space of these KPMs $\F$, and, if we simplify to fixing $M$ and consider the theory in vacuo, i.e. setting $\psi=0$, then $\F=\mathrm{Lor}(M)$, the space of Lorentzian metrics over $M$.\footnote{Indices $a, b, c,$ etc are taken to be \emph{abstract} (cf. \cite[Ch. 2.4]{Wald_book} for an explanation), i.e. only denote the rank of the tensor, but no coordinate basis. I will denote coordinate indices by Greek letters: $\mu, \nu$, etc. }

The physical interpretation of the theory is chronogeometric, in the following sense. According to the \emph{geodesic principle}, the images of smooth geodesic, time-like curves  represent  the possible
histories of freely falling (i.e. subject only to gravity, but to no other force, e.g. electromagnetism) massive test particles. That is, curves $\gamma(\lambda)$,  where $\lambda\in [0,1]$ is some parametrization of the curve,  such that tangent vectors $\gamma'$ are time-like, i.e. $g_{ab}\gamma'^a\gamma'^b <0$, and such that $\gamma'^a\nabla_a\gamma'^b=0$,  represent freely-falling particles whose energy-momentum tensor is ignorable---it doesn't back-react on the geometry.   Those time-like curves that are not geodesic, i.e. do not satisfy $\gamma'^a\nabla_a\gamma'^b=0$,  represent  the possible
histories of massive test particles that are subject to a force additional to gravity, e.g. electromagnetism. Finally, the images of smooth null geodesic curves represent   the possible histories of 
light rays.

In terms of the category-theoretic language (see footnote \ref{ftnt:category}): the groupoid of smooth manifolds has as objects the smooth manifolds, and  diffeomorphisms as the isomorphisms; diffeomorphisms are those maps that preserve the smooth global structure of manifolds. And the category $\mathrm{Lor}(M)$ has as objects the metrics on $M$ of Lorentzian signature $g_{ab}$, and isometries as the isomorphisms.  

The matter and gravitational fields are maps from points of the manifold to some other value space; we will look at this definition in detail when we discuss vector bundles in the second paper, \cite{Samediff_1b}. The dependence of the fields on spacetime points implies that an action by a diffeomorphism $f:M\rightarrow M$ on this base set will lift to an action on the fields: just take the new field to have at $x$ the value that the original field had at $f(x)$. 
We can represent such an action of the diffeomorphisms  of $M$ on a model, represented by the triplet $\langle M, g_{ab},  \psi\rangle $,   by the pull-backs, $\langle M, f^*g_{ab}, f^*\psi\rangle$. 

It is also useful to represent the local, infinitesimal action of diffeomorphisms. Namely, for a one-parameter family of diffeomorphisms $f_t\in \Diff(M)$,  such that $f_0=\mbox{Id}$, the tangent to $f_t$ at $t=0$ is  the vector field $X^a$ ($f_t$ is the flow of $X$). Then, infinitesimally we obtain, for example, for the metric: 
\be\label{eq:Lie_g}\left.\frac{d}{dt}\right|_{t=0} f^*_tg_{ab}\equiv\mathcal{L}_Xg_{ab}=\nabla_{(a}X_{b)}, 
\ee
where $\mathcal{L}_X$ denotes the Lie derivative along $X^a$.\footnote{For a map $f:M\rightarrow N$, for  $\eta$ a one-form on $N$,  $f^*\eta$ is a one-form acting on $v\in T_x M$ as $f^*\eta(v):=\eta (Tf(v))$, where $Tf:TM\rightarrow TN$ is also called the \emph{push-forward} of the map (taking tangents to curves in $M$ to the tangents to the images of those curves under $f$), and is sometimes denoted by a $f_*$. For a scalar function $\rho$ on $N$, and $x\in M$,  $f^*\rho(x)=\rho(f(x))$.  Since, when $f\in \Diff(M)$,  maps and their inverses  are both smooth, we can mostly ignore the distinction between push-forward and  pull-back and denote the appropriate action of the maps without distinguishing superscript and subscript asterisks. Thus, even though formally the pull-back of $f$ would run in the opposite direction of $f$, we will take it to always run in the same direction (by replacing, when necessary, $f$ by its inverse).  \label{ftnt:subs_sups}}   It is useful to denote the diffeomorphisms that are connected to the identity, i.e. that are generated by vector fields through exponentiation, as $\Diff_o(M)$. Here and in the following papers we will mostly focus on this group, as opposed to the full one; but this focus will only be justified in \cite{Samediff_1b}, where it will become important.

If we assume a vacuum, i.e. that $\psi=0$, what are the `natural' isomorphisms of  $\langle M, g_{ab}\rangle $?  
Standard mathematical practice takes isomorphisms in this category to be just those induced by the diffeomorphisms of the base set $M$. 
  Then, \emph{in vacuo}, 
 two models $\langle M, g_{ab}\rangle$ and $\langle M, \tilde g_{ab}\rangle$ are isomorphic if and only if there is a diffeomorphism of $M$, $f\in\,$Diff$(M)$, such that $f(g_{ab})=\tilde g_{ab}$. If matter and radiation fields are included, an isomorphism would require the same map  to similarly relate their distributions in the two models, but these fields could have other isomorphisms beyond those induced by the diffeomorphisms---as we will see.

Thus we have described the isomorphisms of this space of KPMs of vacuum spacetimes. Spacetime physical theories usually assume that these  isomorphisms are also symmetries of the theory, in the sense that a large, salient set of quantities, and their values, will be physically represented equally well by any isomorphism-related model. But what are the dynamical symmetries of the theory? 

In the spirit of Definition \ref{def:inf_sym}, we endow $\F$ with a (infinite-dimensional) manifold-like structure of its own, and define an action functional on this space: $S:\F\rightarrow \RR$, given by:
\be S[g]:=\int_M \d ^4 x\,\sqrt{g}\,R,
\ee
where $R$ is the Ricci scalar curvature of the metric, obtained  by taking the trace of the Ricci curvature, $R:=g_{ab}R^{ab}$.
 We then extremize $S[g]$  in vacuum, and for a fixed boundary-less manifold $M$, so that elements of $\F$ differ only by their metrics.  Then, omitting  indices, from the extremization requirement $S[g+\delta g]-S[g]=0$ for all directions $\delta g\in T_{g}\F$, the equations of motion emerge as conditions on the `base metric' $g$. Besides,  certain vector fields on $\F$ leave $S$ invariant,  e.g. in vacuum $S[g+\widehat{\delta g}(g)]-S[g]=0$, for all $g$, where $\widehat{\delta g}:\F\rightarrow T\F$ is a smooth vector field on this infinite-dimensional field space, $\F$. With another set of minimal assumptions, namely, that $M$ has no boundaries and that $\widehat{\delta g}$ is `local' in a sense to be established below,  these vector fields can be identified as the  the infinitesimal versions of the maps $g_{ab}\rightarrow f^*g_{ab}$. Indeed, these directions can be proven to be given by $\widehat{\delta g}=\mathcal{L}_Xg_{ab}$ of \eqref{eq:Lie_g}, and they generate the isomorphisms induced by the diffeomorphisms of $M$.\footnote{If $M$ has boundaries, then not all vector fields will preserve the value of the action. In that way, there is a departure from dynamical symmetries seen as sets of transformations of the equations of motion which keep `its form' invariant. For if one model satisfies the Einstein equations, an isomorphic model will also satisfy them, irrespective of its behavior at the boundaries. And the addition of a scalar field, $\psi$, would similarly have an infinitesimal symmetry given by $\mathcal{L}_X \psi$.  } 
 
  Here is a sketch of the proof: let $v_{ab}=\widehat{\delta g}$ be a local vector field in $\F$, which  we assume to depend locally on the metric and its derivatives. The local assumption amounts to a definition:\footnote{Just as we would write a general vector field on $M$ as $\sum_i v^i(x)\frac{\partial}{\partial x^i}$. Here the points of $\F$ are $g_{ab}$---in analogy to the points $x\in M$---and suspending the standard Einstein summation convention by reintroducing the explicit summation, we have a more direct analogy with the integration sign in the infinite-dimensional case.  } 
  \be \widehat{\delta g}:=\int_M v_{ab}(g, \pp g,\pp^2 g, \cdots)(x)\frac{\delta}{\delta g_{ab}(x)}.\ee
  We then have a corresponding directional variation of the action functional,
  \begin{align} \delta_vS[g]&=\int_M\d^4\, x\int_M\d^4\, y\,v_{ab}(y)\frac{\delta (\sqrt{g}\,R(x))}{\delta g_{ab}(y)}\\
  &=\int _M \d^4\, x\,\sqrt{g}\left(\nabla_a\nabla_bv^{ab}-\nabla^c\nabla_c (g^{ab}v_{ab})-v_{ab}R^{ab}-\frac12 R g_{ab}v^{ab}\right) \\
  &\hat = \int_M\d^4\, x\,\sqrt{g} \, v_{ab}(R^{ab}-\frac12 g^{ab}R)\,\overset{!}{ =}\,0,\label{eq:deltaS}
  \end{align}
 where $\hat =$ is equality up to boundary terms (which we are taking to vanish) and $\overset{!}{=}$ uses our assumption of symmetry. Since equality up to boundary terms must hold at all $g_{ab}$, and thus for all $R_{ab}$ (not just those that satisfy the Einstein equations), it is not hard to see that the only way to ensure \eqref{eq:deltaS} is to make use of the general geometric constraints on $R^{ab}-\frac12 g^{ab}R$: namely,  the algebraic symmetry of the indices $ab$ and the contracted Bianchi identity, $\nabla_a(R^{ab}-\frac12 g^{ab}R)=0$. Since $v_{ab}$ is already symmetric in $ab$,\footnote{Because $\frac{\delta}{\delta g_{ab}(x)}$ is symmetric in $ab$, and for any two tensors, $Y^{ab}Z_{(ab)}=Y^{(ab)}Z_{(ab)}$.} there is no further use for the algebraic symmetry; we can only profitably use the Bianchi identity. Since the Bianchi identity involves contraction of the term multiplying $v_{ab}$ with a covariant derivative, we must have at least one total derivative inside $v_{ab}$, and we can then use integration by parts  (using integration by parts leaves only a boundary term, which vanishes by assumption). Therefore, the only completely general solution is to take $v_{ab}=\nabla_{(a} X_{b)}={\cal L}_X g_{ab}$. Note that this argument works for completely general, possibly metric-dependent, vector fields $X^a\in \mathfrak{X}(M)$. 
 
  Thus, following Definition \ref{def:inf_sym}, we obtain the full set of symmetries of the theory; already a remarkable triumph of the definition. In contrast,  as far as I know:  without using the infinitesimal definition and applying it to the action functional, the  proof that the most general symmetries of general relativity were given by generalized diffeomorphisms (and constant dilatations) was only provided relatively recently,  in  \cite{Torre_GRsym}.\footnote{More rigorously, the proof of uniqueness of the solution here would proceed in much the same way as Lovelock's theorem (1973), which is nonetheless much simpler than \cite{Torre_GRsym}'s proof. There, they prove that the only infinitesimal, generalized symmetries of the equations of motion of general relativity, i.e. general, metric dependent transformations of the Einstein tensor that would vanish when the original Einstein tensor vanishes, are $v_{ab}=\nabla_{(a} X_{b)}$ and $v_{ab}=c g_{ab}$, where $c$ is a constant. Here, the constant dilatation does not emerge, because our symmetries of the action must also hold when the Einstein tensor doesn't vanish (i.e. also hold off-shell). }  
 
 And of course, these directions in $\F$ are integrable, forming a closed space,  since the Lie derivative obeys $\mathcal{L}_X\mathcal{L}_Y- \mathcal{L}_Y\mathcal{L}_X=\mathcal{L}_{[X, Y]}$, where $[\bullet, \bullet]$ is the commutator of vector fields. 
 So these infinitesimal symmetries, by \eqref{eq:Lie_g}, generate diffeomorphisms, and diffeomorphisms form orbits of Lor$(M)$. 
Thus we identify the symmetry group as $\G:=\Diff_o(M)$, which acts pointwise on the space of Lorentzian metrics over $M$, namely, $\F=\mathrm{Lor}(M)$.

Therefore, in vacuo, we will say that $\langle M, g_{ab}\rangle$ and $\langle M, \tilde g_{ab}\rangle$ are both isomorphic and symmetry-related  iff there is an $f\in \Diff_o(M)$, such that $\tilde g_{ab}=f^*g_{ab}$. We write this as: 
\be\label{eq:equivalence}
\langle M, g_{ab}\rangle\sim \langle M, f^*g_{ab}\rangle.
\ee

Before we turn to gauge theories, I would like to emphasize two points. First, 
I find it remarkable that such a general definition as Definition \ref{def:inf_sym}, when applied to the action functional, already implies so much structure for symmetry, such as being integrable into an orbit, and having group structure.\footnote{In this respect, the covariant symplectic formalism is very convenient: we can define symmetries as vector fields in the kernel of the symplectic form; then a few steps suffice to show that these vector fields form an algebra that lies  also in the kernel of the symplectic form, and thus through exponentiation we obtain  the orbits of the symmetry group (cf. \cite[Sec. 2]{Lee:1990nz}). Indeed, the null directions of   $\Omega$ are necessary and sufficient to characterise the generators of gauge symmetry. For suppose the vector fields $v, w$  are such that $\Omega(v, \bullet)=0=\Omega(w, \bullet)$.  Using the Cartan Magic formula relating Lie derivatives, contractions $i$ and the exterior derivative $\d $:   
$$
 \mathcal{L}_{v}\Omega=(\d i_{v}+ i_{v}\d)\Omega=0; 
$$
i.e. the first term  vanishes because $\Omega(v,\bullet)=0$ and second because $\d \Omega=0$. So $\Omega$ itself is invariant along $v$. Moreover, if we take the commutator of $v, w$, i.e. $[ v, w]= \mathcal{L}_{v}w$,  contract it with $\Omega$, and remember the formula:
$$\mathcal{L}_{v}(\Omega(w, \bullet))=\Omega( \mathcal{L}_{v}w, \bullet)+  (\mathcal{L}_{v}\Omega)(w, \bullet) \, ,
$$
we obtain that, since both $\mathcal{L}_{v}(\Omega(w, \bullet))=0$  and $\mathcal{L}_{v}\Omega=0$, it is also the case that $\Omega( [{v}, w], \bullet)=0$. Thus, by the Frobenius theorem  the kernel of $\Omega$ forms an integrable distribution which  integrates to give  the orbits of the symmetry transformation. 

 But it is important to stress that while the covariant Lagrangian version of both Yang-Mills theories and general relativity have groups of symmetries, and so does the Hamiltonian version  of Yang-Mills theory  (in which ${\F}$ is phase space), the set of symmetries of the Hamiltonian version of general relativity has only a groupoid structure (see \cite{Blohmann_2013}).}

The second point to note is that diffeomorphisms act transitively \textit{on} $M$; any point can be carried to any other point. This  means that there is no non-trivial orbit for $\Diff_o(M)$ picking out subsets of $M$. Of course, $\Diff_o(M)$ does \emph{not} act transitively on  the infinite-dimensional $\mathrm{Lor}(M)$: the orbits of $\G$ by \eqref{eq:equivalence} are closed subsets of that domain, and are said to \emph{foliate it}. 
Thus diffeomorphisms and gauge-symmetries are indiscernible \emph{at the level of $\F$}; to discern them---as we will more completely do more completely in \cite{Samediff_2}---we must zoom in on their action on the base manifold $M$, or what we will call the \emph{pointwise} action of the symmetries.


\section{Gauge transformations in Yang-Mills  theories}\label{sec:PFB}
This Section will explore details of symmetries in gauge theories: more especifically, of Yang-Mills theories.

Speaking metaphysically, the  previous Section \ref{sec:attitudes} construed the symmetries of general relativity as isomorphisms of a natural geometric structure. And there is a possible  misgiving that the symmetries of gauge theory are less natural, and thus have a less natural structural interpretation than those of general relativity. 

I believe that the concern is indeed justified in the  case of gauge transformations in the gauge-potential formalism for electromagnetism; I will explain this in  Section  \ref{sec:crude_A}, after I have described the basics of that formalism in Section \ref{sec:basic_EM}.  But that formalism is not the last word in the theoretical development of Yang-Mills theories. In  Section  \ref{sec:PFB_mot} I motivate the need for  a more complete,  geometric understanding of what the fields and gauge symmetries of modern physics are about. We leave a brief presentation of the  mathematical formalism to Section \ref{sec:PFB_formalism}. 


\subsection{Electromagnetism in the gauge potential formalism: basics}\label{sec:basic_EM}

In electromagnetism, the basic dynamical variable is the electromagnetic field tensor, $F_{ab}$. Upon choosing a spacetime split into spatial and time directions, the components of the electromagnetic tensor become the familiar electric and magnetic fields (in coordinates): $F_{i0}=E_i$, and $F_{jk}\epsilon_i^{jk}=B_i$  (where we used the three-dimensional totally-antisymmetric tensor, $\epsilon$, or the spatial Hodge star, to obtain a 1-form). 

The Maxwell equations in the Minkowski spacetime are written, in a coordinate basis, in terms of $F_{{a}{b}}$, as:
\be\label{eq:EM}\nabla^{a} F_{{a}{b}}=j_{b}, \quad \text{and}\quad \nabla_{[{a}}F_{{b}\kappa]}=0, \ee
where   $j_a$ is the current, and square brackets denote anti-symmetrization of indices. The second equation of \eqref{eq:EM} is called `the Bianchi identity', and it is read as a  constraint on the field tensor. A geometric explanation for this constraint is that $F_{{a}{b}}=\nabla_{[{a}}A_{{b}]}$, or, in exterior calculus notation, $\d \mathbf{A}=\mathbf{F}$, where $A_{a}$ is called \textit{the gauge-potential}. At least locally, this relation follows from the Poincar\'e lemma. 

The equations of motion of this theory---now assuming \emph{in vacuo}, i.e. $j=0$, for simplicity---are: 
 \be\label{eq:eom_A} \nabla^{a}\nabla_{b} A_{a}-\nabla^{a}\nabla_{a} A_{b}=0.
 \ee
 And these equations are obtained from the action functional:
  \be\label{eq:action_A}S[A]:=\int_M \nabla_{[{a}}A_{{b}]}\nabla^{[{a}}A^{{b}]}=\int_M *\mathbf{F}\wedge \mathbf{F},
 \ee
 where $*$ is the Hodge-star operator (which takes an argument differential form to its ortho-complement, and $\wedge$ is the exterior (wedge) product between forms). 
 
 The classical interpretation of the theory interprets the Faraday tensor as a physical field, e.g. the electric and the magnetic, in a given space-time split. If we add to the Lagrangian the contribution from a charged particle with charge $q$ and mass $m$, whose  world-line is given by $\gamma$: 
 \be S_{\text{\tiny{particle}}}=\int_\gamma (m \gamma'^a\gamma'^b g_{ab}+q A^a \gamma_a),
 \ee
 we obtain, from the variation with respect to the particle trajectory, the Lorentz-force law: 
 \be m \gamma'^a\nabla_a\gamma'_b=q F_{ab}\gamma'^a, 
 \ee
 which describes how the motion of charged particles is disturbed by electromagnetic interactions. 
 
 Non-Abelian Yang-Mills theories have an analogous equation of motion, \eqref{eq:non_Ab} below, which, like the Einstein equations, are only gauge-\emph{co}variant, and not gauge-\emph{in}variant like the Abelian version of electromagnetism; and they have an analogous Lorentz-force equation as well. But due to quantum effects---namely, \emph{confinement}---the theory does not have a long-ranged classical interpretation like electromagnetism does. 
\subsection{Symmetries need not be isomorphisms: an example from gauge theory}\label{sec:crude_A}

 Gauge-potentials for electromagnetism are locally just smooth one-forms on the manifold, and the natural notion of isomorphism here is just the one inherited from differential geometry: again, pull-backs by spacetime diffeomorphisms. That is, the KPMs of the theory  are given by  $\langle M, \mathbf{A}\rangle$, where $\mathbf{A}$ is given by $A_a$, or, in coordinates, $A_\mu \d x^\mu$, i.e. the potentials are  sections of the cotangent bundle---real-valued one-forms over each topologically trivial patch---on the manifold $M$.  Since they are differential forms, we could rehearse the argument of Section  \ref{sec:diff_sym} and conclude that  the isomorphisms of the space of models are again pull-backs by diffeomorphisms. 
 
 But the dynamics of the theory are another matter.  
 If we  follow the definition of symmetries given in Section  \ref{sec:syms_tech}, we arrive at the standard gauge transformations. 
 
 Namely, in analogy to \eqref{eq:deltaS}, we have here:
  \be\label{eq:syms_EM} \delta_v S[A]= \int_M   F^{{a}{b}}\nabla_{b} v_{a}\overset{!}{=}0.
  \ee 
  Now, we are not allowed to use the equations of motion, since this equality must hold for general $A_a$. Here, the only general constraint at our disposal to solve \eqref{eq:syms_EM} is the algebraic anti-symmetry of $ab$.\footnote{Why can't we use the Bianchi identity, once again? Because here, the indices are already contracted, i.e. after integration by parts we obtain  $\nabla_aF^{{a}{b}}$. In form language, there is no local operator that, acting on $\d *\mathbf{F}$, will result in something proportional to $\d \mathbf{F}$ (which is what vanishes due to the Bianchi identity).} Thus we must have that $\nabla_{[b} v_{a]}=0$. As a one-form, we rewrite this as  $\d\mathbf{v}=0$, which, by the Poincar\'e lemma, implies that locally  $\mathbf{v}=\d \xi$, for a scalar function $\xi$. 
Thus the infinitesimal symmetry adds a gradient of a smooth function to the gauge-potential one-form: $\mathbf{A}\rightarrow \mathbf{A}+\d \xi$, for $\xi\in C^\infty(M)$.

Here, there is no analogue of \eqref{eq:Lie_g} for the symmetries: no isomorphism of an underlying space induces the symmetry through pull-back. The dynamical symmetries are therefore `larger' than those expected from the natural notion of mathematical isomorphisms of the objects in play, which would, again, be diffeomorphisms.\footnote{Here the natural  symmetries involve only differential geometric operations---such as exterior differentiation---and thus composition with diffeomorphisms is well-defined. Indeed, the two operations  commute, since the exterior derivative commutes with the pull-back: for $f\in$Diff$(M)$,  the object and arrow $(\mathbf{A}, \xi)$ gets mapped to $(f^*\mathbf{A}, f^*\xi)$. \label{ftnt:comm_iso}} 

Nonetheless, there are natural geometric structures for which gauge transformations emerge as isomorphisms. We will look at these geometric structures in the upcoming Section \ref{sec:PFB_mot}. But we will only fully justify the correspondence between local gauge transformations and the automorphisms of this structure in Section \ref{sec:p-a}.  

\subsection{A brief introduction to fiber bundles}\label{sec:PFB_mot}

The modern mathematical formalism of gauge theories relies on the theory of principal  and associated fibre bundles. We will not give a comprehensive account here (cf. e.g. \citep{kobayashivol1}).  In this Section  we  introduce the necessary ideas, and in Section \ref{sec:PFB_formalism} we introduce the formalism in more detail.

Our  intuitive idea of a field over space is something like temperature. A temperature field can be written as a map from space to the real numbers, $T:M\rightarrow \RR$. Being told that there are fields that have a more complicated `internal structure' than temperature---for instance, vector fields that over each point of spacetime can point in different directions---we will want to generalize a  scalar map like temperature  to  $\rho:M\rightarrow F$, a map from spacetime to some internal vector space $F$. 

For tensor bundles, made up of tensor products of tangent and cotangent vectors, $F$ is ``soldered'' onto spacetime, $M$.\footnote{For instance, we can identify elements of the tangent bundle with tangent vectors of curves on the base manifold. In more detail, supposing  the internal vector space $F$ has the dimension of $M$, a \emph{soldering form} gives an isomorphism between each $T_xM$ and $F$, in a smooth way. }
 But the fields employed in modern theoretical physics---representing different properties of matter---live in more general vector bundles, $F$, which are not thus soldered to spacetime. Generically,  those fields   have many components at each spacetime point which are not associated to spacetime directions; they represent  degrees of freedom that are `internal' to each spacetime point. 
 Such fields interact through  forces other than the gravitational force, and each of these forces is related to a given gauge or symmetry group, because certain properties of these interactions reflect some symmetry group.  
 
 The worry might arise that the same  symmetry group could be realised very independently on different  matter fields.  But  all these different matter fields  interact  with the same force, and thus  the action of the symmetry group must be meshing  between the various matter  fields. Mathematically, this  means that the parallel transport of internal quantities is compatible for all the fields. This `coincidence' is conveniently described if we encode the symmetries through the formalism of \emph{principal fiber bundles} (PFBs): they allow us to encode the essential symmetry structure of each type of interaction---e.g. electromagnetic---independently of the individual matter fields that are susceptible to this interaction.  



 In more detail, states of different species of matter are represented in (as sections of) different vector bundles: one vector bundle per field. The main idea of a principal fiber bundle is that it is a space where a given Lie group---usually taken to be associated with a certain type of fundamental force or interaction---acts.  And then, as expounded lucidly by \cite{Weatherall2016_YMGR},   the \emph{Ehresmann connection} of a principal bundle regiments  the symmetry properties of all the different  matter  fields that feel that force or interaction.  Charged scalar, electron, quark-fields, etc., all interact electromagnetically; and that interaction is mediated by the same fundamental electromagnetic field (\emph{mutatis mutandis}, for other interactions, e.g. replacing  `electromagnetism' by the `strong force'). This means that the relevant covariant
derivative operators on the vector bundles in which these matter fields are valued  have the same
parallel transport and curvature properties.   Such  universality is mathematically enforced because these vector bundles are associated to  the same fundamental Ehresmann connection on the principal bundle, and this means they have their covariant derivative operators defined uniquely by that connection. 

  In this respect, the role performed by the geometry of spacetime in mediating  gravitational interactions is precisely analogous to  the role performed by  the geometry of the principal bundle in mediating other forces or interactions. In a direct analogy: just as the Levi-Civita connection in gravity encodes the geometric properties of the gravitational force and dictates the parallel transport of fields that interact gravitationally, the Ehresmann connection encodes the geometric properties of some other force,  and dictates parallel transport of components of the fields that interact with that force.  

The main idea underlying the physical significance of the parallel transport of internal quantities was already well stated  in the  paper that introduced this mathematical machinery into physics,   \citet{YangMills}: 
\begin{quote} The conservation of isotopic spin is identical with the requirement of invariance of all interactions under
isotopic spin rotation. This means that when electromagnetic interactions can be neglected, as we shall hereafter assume to be the case, the orientation of the
isotopic spin is of no physical significance. The differentiation between a neutron and a proton is then a
purely arbitrary process. As usually conceived, however,
this arbitrariness is subject to the following limitation:
once one chooses what to call a proton, what a neutron,
at one space-time point, one is then not free to make any
choices at other space-time points. \end{quote}
The idea here is that calling a particle a proton or a neutron at a given point is meaningless; only  \textit{relational} or, more broadly,  \textit{structural} properties of the theory can have physical significance, for instance, whether your original `proton' became a `neutron' upon going around a loop.\footnote{ Of course this example, which originally motivated Yang and Mills, applies only in the context of  isospin symmetry---which is approximate. For the electric charge tells protons and neutron apart in an intrinsic manner.} 
The only physically relevant information is a notion of  sameness across different points of spacetime: thus, once  we label a given particle as e.g. a proton at one point of spacetime, the structure of the bundle specifies what would also count as a proton at another spacetime point, infinitesimally nearby. 
These constraints are imposed by the Ehresmann connection-form, or connection-form for short. 
A connection-form $\omega$ maps infinitesimally nearby points of the manifold $P$ (on which the group acts) to infinitesimal group elements. In Section \ref{sec:PFB_formalism}, we give the technical conditions that make precise this idea.  


 \subsection{The formalism of principal fibre bundles}\label{sec:PFB_formalism}\label{subsec:general_cons}


  A principal fibre bundle is, in short, just a manifold where some group acts. In detail: it is a smooth manifold $P$ that admits a smooth free action of a  (path-connected, semi-simple) Lie group, $G$: i.e.  there is a map $G\times P\rightarrow P$ with $(g,p)\mapsto g\cdot p$ for some left action $\cdot$ and such that for each $p\in{P}$, the isotropy group is the identity (i.e. $G_p:=\{g\in{G} ~|~ g\cdot p=p\}=\{e\}$). 
  
Naturally, we construct a projection  $\pi:P\rightarrow{M}$ onto equivalence classes, given by  $p\sim{q}\Leftrightarrow{p=g\cdot{q}}$ for some $g\in{G}$. That is: the base space $M$ is the orbit space of $P$, $M=P/G$, with the quotient topology, i.e. it is characterized by an open and continuous $\pi:P\rightarrow M$. By definition, $G$ acts transitively on each fibre, i.e. orbit. The automorphism group of $P$---those transformations that preserve the  structures---are \emph{fiber-preserving} diffeomorphisms, i.e. diffeomorphisms
\be\label{eq:tau}\tau:P\rightarrow P \quad \text{such that}\quad  \tau(g\cdot p) =g\cdot \tau(p).\ee 
Purely internal, or gauge transformations can be identified as those for which $\pi\circ\tau\circ\pi^{-1}=\mathrm{Id}_M$; that is,  as purely `vertical' automorphisms of the bundle; (the orbits are usually drawn going up the page, hence `vertical').

\subsubsection{ The Ehresmann connection-form.}
On $P$, we consider an Ehresmann connection $\omega$, which is a 1-form on $P$ valued in the Lie algebra $\mathfrak{g}$ of $G$ that satisfies appropriate compatibility properties with respect to the fibre structure and the group action of $G$ on $P$. 
The connection selects a ``vertical'' subspace of the tangent space $T_pP$ at $p\in P$, which ``points in the direction of the fiber'', and  it selects a ``horizontal'' subspace---which gives the notion of parallel transport linking nearby fibres. 
 Namely: 
Given an element $\xi$ of the Lie-algebra $\mathfrak{g}$, we define the vertical space $V_p$ at a point $p\in P$, as the linear span of vectors of the form 
\be\label{eq:fund_vec}v_{\xi}(p):=\frac{d}{dt}{}_{|t=0}(\exp(t\xi)\cdot p), \quad \text{for}\quad \xi\in \mathfrak{g}.\ee
 And then the conditions on $\omega$ are:
\be\label{eq:omega_defs}
\omega(v_\xi)=\xi
\qquad\text{and}\qquad
{L_g}^*\omega=g^{-1}\omega g,
\ee
  where ${L_g}^*\omega_p(v)=\omega_{g\cdot p}({L_g}_* v)$ and where ${L_g}_*$ is the push-forward of the tangent space for the left-action $g:P\rightarrow P$. Thus, we can only characterize the action of $\omega$ on vector \textit{fields} on $P$, i.e. on sections of the vector bundle $TP$,  say $\zeta\in C^\infty(TP)$, if they are \textit{left-invariant}, i.e. if $\zeta_{g\cdot p}={L_g}_*\zeta_p$.
  
  A choice of connection is equivalent to a choice of covariant `horizontal' complements to the vertical spaces, i.e. $H_p\oplus V_p=T_pP$, with $H$ compatible with the group action.   That is, since $\omega$ is $\mathfrak{g}$-valued and gives an isomorphism between  $V_p$ and $\mathfrak{g}$, the first condition of \eqref{eq:omega_defs} means that: i) the kernel $\mathsf{Ker}(\omega_p)=H_p$, and ii)  since $V_p=\mathsf{Ker}(\pi_*)$, $H_p$ will be 1-1 projected by $\pi_*$ onto the tangent space $T_{\pi(p)}M$. Thus the vectors spanning $\mathsf{Ker}(\omega_p)$ are the so-called \textit{horizontal} vectors in the bundle, and each represents a unique `horizontal lift' at $p$ of a direction at $T_{\pi(p)}M$. This condition also requires that, much like the metric, the connection form is nowhere vanishing. The second condition of \eqref{eq:omega_defs} guarantees that the notion of horizontality covaries with the choice of representative of the fiber (e.g. the choice of frame in the frame bundle example above), that is: a vector $v\in T_pP$ is horizontal iff ${L_g}_* v\in T_{g\cdot p}P$ is horizontal.

 To define curvature, we note that an infinitesimally small parallelogram with horizontal sides that projects onto a closed parallellogram on $M$, may not close on $P$. Namely, if a horizontal parallellogram starts at $p\in P$, it may end at $g\cdot p$. Infinitesimally, we obtain a Lie-algebra valued two-form on $P$, 
 \be \Omega:= \d \omega+\omega\wedge\omega,\ee
  where $\d$ is the exterior derivative on $P$. 
  

 
 \subsubsection{The gauge  potentials.}\label{sec:gauge_pot}

Locally over $M$,  it is possible to choose a smooth embedding $\sigma$ of the group identity into  the fibres of $P$. Called  {\it local sections} of $P$, these  are maps $\sigma:U \to P$ such that $\pi\circ \sigma = \mathrm{id}$. So for $U\subset M$, there is a map ${\sigma}: U\rightarrow P$ such that $P$ is locally of the form $U\times G$.

Given  local sections $\sigma$ on each chart domain $U$, we define a local spacetime representative $\mathbf{A}$ of $\omega$,  as the pullback of the connection, $\mathbf{A}^\sigma:=\sigma^*\omega \in \Lambda^1(U_\alpha, \mathfrak{g})$; (here  $\sigma$ is \textit{not} a spacetime index; we momentarily keep it in the notation as a reminder of the reliance on a choice of section).\footnote{Note that $\mathbf{A}$ only captures the content of $\omega$ in directions that lie along the section $\sigma$. The vertical component of $\omega$---which is dynamically inert, as per the first equation of \eqref{eq:omega_defs}---can be seen (in a suitable interpretation of differential forms, cf. \cite{Bonora1983}) as the BRST ghosts. This interpretation   geometrically encodes gauge transformations through the BRST differential \cite{Thierry-MiegJMP}. Although interesting in its own right, we will not explore this topic here. See \cite{GomesStudies, GomesRiello2016} for more about the relationship between ghosts and the gluing of regions. \label{ftnt:ghosts}} Similarly, we can define the field-strength $\mathbf{F}^\sigma=\sigma^*\Omega$. We will expand on the physical  significance of these sections in \cite{Samediff_1b}, and in Section . 

In a basis for a given chart on $U\subset M$, we write: $\mathbf{A}=A_\mu^I \,\d x^\mu \tau_I,\,\, \tau_I\in \mathfrak{g}$ is a Lie-algebra basis,  and $\, A_\mu^I \in C^\infty(U)$.\footnote{Clearly, $I$ are Lie-algebra indices and $\mu$ are spacetime indices. We take $\{\d x\otimes \tau\}$ to stand in for  the basis for a vector bundle $T^*U\otimes \mathfrak{g}$.} Vertical automorphisms are represented as gauge transformations, which, infinitesimally, for a Lie-algebra valued function $\xi^a\in C^\infty(U, \mathfrak{g})$,  act as 
\be\label{eq:gauge_trans}\delta_\xi A_\mu^I= \pp_\mu\xi^I+[ A_\mu, \xi]^I=\D_\mu\xi^I,
\ee
where $\D_\mu(\bullet)=\pp_\mu(\bullet)+[A_\mu, \bullet]$, the gauge-covariant derivative, is defined to act on Lie-algebra valued functions. 

The equations of motion of Yang-Mills theory, written as differential equations of fields on spacetime, are: 
\be\label{eq:non_Ab} \D^\mu F_{\mu\nu}^I=j_\nu^I, 
\ee
where $F_{\mu\nu}^I=\nabla_{[\mu}A_{\nu]}^I-\frac12[A_\mu, A_\nu]^I$, and $j_\nu^I$ is the charged non-Abelian current. 

 \subsubsection{PFBs as bundles of linear bases and associated bundles}\label{subsec:PFB_ex}

To gather intuition about principal fiber bundles (PFBs)  as the `organizers' of symmetry principles, as described in Section \ref{sec:PFB_mot}, it is worthwhile to introduce them in the context of the familiar tangent vector fields on $M$. 

 The main idea of fiber bundles is that they are spaces that locally look like a product, i.e. a fiber `bundle'. So the many fields of nature would be represented as maps that take each point of spacetime (or space)  into its respective value space, or fiber. 

We denote fiber bundles by $E$; they are smooth manifolds that admit the action of a surjective projection $\pi:E\rightarrow M$ so that locally $E$ is of the form $\pi^{-1}(U)\simeq U\times F$, for  $U\subset M$ (and similarly for all subsets of $U$) and $F$ is some `fiber': a  space that `inhabits' each point of $M$ and in which the fields take their values. 

 But the decomposition $\pi^{-1}(U)\simeq U\times F$ is not unique, and will depend on what is called `a trivialization' of the bundle, which is basically a coordinate system that makes the local product structure explicit.  Thus, in principle there is no unique identification of an element of $F$ at a point $x\in M$ with an element of $F$ at a point $y\in M$. In principle, there is no identification of a vector, or even of a scalar quantity, like temperature, as possessed at different points of spacetime.

So, to be explicit: $F$ is some space where we can have quantities in spacetime take their value; for instance,  a scalar field could take values in $\RR$ or $\bb C$, whereas a more complicated field such as a  vector  field or a spinor field, could take values in $\RR^4, \bb C^4$, etc. A choice of \textit{section} of the bundle  represents fields taking values in $F$: e.g. a spinor field, or a quark field, etc, which are all vector bundles, in that $F$ is a vector space. A field-configuration for $E$ is called \textit{a section}, and it is a map $\kappa: M\rightarrow E$ such that $\pi\circ\kappa=\mathrm{Id}_M$.\footnote{ It is somewhat confusing that a \textit{section of a vector bundle} is an entirely different object from the section of a principal bundle. So, for instance two different choices of the electron field are two different sections of its vector bundle, and thus are not counted as `equivalent' in the way that two sections of a principal bundle are. And while a global section of $P$ exists iff the bundle is trivial, we can always find a global section of an associated bundle (cf. \cite[Theo. 5.7]{kobayashivol1}).\label{sec:section_triv}}  Sections replace the functions $\tilde\kappa:M\rightarrow F$, that we would employ if the fields that physics uses had a fixed, or  ``absolute''---i.e. spacetime independent---value space. We denote smooth sections like this by $\kappa\in C^\infty(E)$.

 A useful example of a vector bundle is  the tangent bundle, $TM$. A  smooth tangent vector field is  a smooth assignment of  elements of $TM$ over $M$,  denoted  $X\in \mathfrak{X}(M)$, with $\pi:TM\rightarrow M$, mapping $X\in T_xM\rightarrow x\in M$. The tangent bundle   $TM$ \emph{locally}  has the form of a product space, $U\times F$, with $F\simeq \RR^4$.  But even if $TM$ were globally trivializable, so that a product structure could be found for its totality,  this would  not mean we could identify an element $v\in \RR^4$ at different points of $M$. Differential geometry teaches us to attach a vector space to each point of $M$ and to have vectors at different points objectively related only according to some definition of parallel transport along paths in $M$.

This example is also useful to articulate what we mean by a principal fiber bundle that `orchestrates the parallel transport' of the other fields. Here the principal bundle that orchestrates parallel transport  of tangent vectors (and tensor bundles in general)  can be taken to  be the bundle of linear frames of $TM$, called `the frame bundle' (where `frame' means `basis of the tangent space $T_xM$'), written $L(TM)$.  The fibre over each point of the base space $M$ consists of all of the linear frames of the tangent space there, i.e. all choices $\{\mathbf{e}_I(x)\}_{I=1, \cdots 4}\in L(TM)$, of sets of spanning and linearly independent vectors (here the index $I$ enumerates the basis elements).\footnote{Depending on the theory, we will take different subsets of the linear frames, and of the corresponding structure group. For instance, for general relativity, we take the structure group as $O(4)$ (or $SO(3,1)$)  acting on the orthonormal bases.}  

So each point $p\in P$ of the frame bundle above a point $x\in M$ (i.e. such that $x=\pi(p)$) is just a  basis for the tangent space $T_xM$; and there is a one-to-one map between the group $GL(\RR^4)$ and the fibre: we can use the group to go from any frame  to any other (at that same point), but there is no basis that canonically corresponds to the identity element of the group.  This example illustrates a feature of principal fiber bundles that distinguishes them from vector bundles: in the former, the fibers are isomorphic to some Lie group $G$; and there is no ``zero'' or identity element on each fibre, as there is in a vector bundle.

 If we imagine the orbits of the group, or the fibers, as being in the vertical direction,  directions transversal to the fiber will connect frames over neighbouring points of $M$.  We thus dub as \textit{horizontal} those directions by which a connection identifies---or `links'  and takes  as identical---frames on neighbouring fibers.\footnote{
In general relativity, we could take this to be  a torsion-free connection-form on $P$ by $\d \mathbf{e}^I=\omega_J^I \mathbf{e}^J$, where $\omega$ here satisfies the expected equations, see \eqref{eq:omega_defs} below (and we used the one-forms  algebraically dual to the vector basis: $\mathbf{e}^J(\mathbf{e}_I)=\delta^J_I$). This equation translates to one using the covariant derivative $\nabla$ as: $\nabla \mathbf{e}_I=\omega^J_I \mathbf{e}_J$.\label{ftnt:gr_viel} } That is: to link fibres, we need to postulate more structure: a connection.

To see how these horizontal directions encode parallel transport of vectors, we need to return to the tangent bundle $TM$, from the frame bundle, $L(TM)$. We proceed as follows:  take a point of $TM$, i.e. a vector at a given point $x\in M$, $X_x\in E_x$ as an element of the fiber $ T_xM\simeq F=\RR^4$, where the ordered quadruplet are the components of $X_x$ according to a frame, $\{\mathbf{e}_I(x)\}\in L(TM)$. So, we write $X_x=a^I \mathbf{e}_I\in T_xM$ as the ordered quadruplet $(a^1, \cdots, a^4)\in \RR^4$. Of course, if we rotate the frame by an element of the group in question, i.e. $GL(\RR^4)$, say by a matrix $g^{IJ}=\rho(g)$, where $\rho:G\rightarrow GL(\RR^4)$ is the matrix representative of the abstract group, then, as long as we undo that rotation on the components, we obtain the same vector, in the original frame. That is, $a^K g_{KL}^{-1} g^{LI}e_I=a^I e_I$. Thus, if we write a doublet $(p,v)$ as, respectively, the frame and the components, we want to identify $(gp, vg^{-1})$ (where we have simplified the notation for the action of the group to be just juxtaposition). This is a standard construction of an \emph{associated bundle}, denoted by $TM\simeq L(TM)\times_\rho \RR^4$. 

Once we have constructed associated bundles in this way, parallel transport, for any vector bundle comes naturally from a notion of horizontality in the principal bundle. To find the parallel transport of the vector $X_x$ along $Y_x$, take the curve $\gamma(t)\in M$ with $\gamma(0)=x$, and so that $\gamma'(0)=Y_x$. Given a frame $p_x\in P$ so that $\pi(p_x)=x$, we take the horizontal lift of $\gamma(t)$ through $p_x$: call it $\tilde \gamma(t)$. Let $X_x=[p_x, v]$, where $v\in \RR^n$ are the components of $X_x$ in terms of the basis $p_x$. 
By definition, the curve in $E$ given by $[\tilde\gamma(t), v]$ is parallel transported, i.e. gives a parallel transport of $X_x$ along $\gamma(t)$. Now, we can define the covariant derivative of a vector field  $X$  such that $X(x)=X_x$ as follows. First, we define $v_X:P\rightarrow \RR^4$   such that, for all $p\in P$
\be\label{eq:vectorP} X(\pi(p))=[p, v_X(p)],\quad \text{where}\quad v_X(g\cdot p)=g^{-1}v_X(p);\ee
that is,  $v_X(p)$ is the decomposition of $X(\pi(p))$ on the basis $p$ (and  therefore $v_X$ obeys the covariance property on the right of \eqref{eq:vectorP}). Thus we define the covariant derivative of $X$ along $Y$ at $x$, as:
\be \nabla_YX(x):=\lim_{t\rightarrow 0}\frac{1}{t}([\tilde\gamma(-t), v_X(\tilde\gamma(-t))]-[p_x, v_X(p_x)]).
\ee
In words, we compare the parallel transported components of $X$ with the actual components of $X$; their non-constancy corresponds to the failure of $X$ to be parallel transported, and to the non-vanishing covariant derivative of $X$.  In this way a covariant derivative is just the standard derivative of the components in the horizontal---or parallel transported---frame. This is, in words, the description of the covariant derivative of $X$ along $Y$ at $x\in M$. 

The picture is useful in that it applies to any vector bundle on which the structure group $G$ in question acts. For instance, in the standard model of particle physics, the fundamental forces are associated to Lie groups, and each field that interacts via such  a force lives in a vector bundle that admits an action of the corresponding group. Thus for a given vector bundle with typical fiber $F$, we have a linear representation of the Lie group in question, $G$, $\rho:G\rightarrow GL(F)$, and we can take the principal connection---the notion of horizontality in the PFB with structure group $G$---to induce a notion of parallel transport in the bundle $E$ with fiber $F$. Indeed, we can take the same procedure as above, building a linear frame for $F$ at each point; parallel-transport then encodes an appropriate $G$-covariant way to identify vector values along paths in the base space $M$. 


\section{The correspondence between active and passive transformations}\label{par:active_passive}\label{sec:p-a}

In the previous Section we saw an interesting contrast: in one formulation of Yang-Mills  theory, the  symmetries are isomorphisms that are induced from the automorphisms of a natural geometric structure---a fibered manifold. In another, the symmetries are just postulated, and, at least on the surface, have nothing to do with the automorphisms of a geometric structure.\footnote{These two formulations of the theory are paradigmatic examples of candidates for \emph{internal} and \emph{external} \emph{sophistication}, respectively, to be studied in \cite[Sec. 4]{Samediff_1a} (see \cite{Dewar2017}).} 

But it is possible to show that the postulated symmetries in fact arise as mere passive transformations---coordinate changes---of the natural geometric structure. Conversely, the maps that change coordinates can also induce a subset of the active automorphisms of the geometric structure. Thus, in this Section I will show that there is an interesting one-to-one correspondence between the two kinds of symmetry, in both the general relativistic and in the Yang-Mills case, at least for the infinitesimal symmetries of Section \ref{sec:syms}. This correspondence is usually understood, in the spacetime case, as one between the active and passive diffeomorphisms. It is this correspondence that  justifies the pragmatic physicists' nearly universal focus on coordinate transformations in lieu of active transformations.

I will start by describing the general relativistic case, in Section \ref{subsec:passive_active_GR}. Then in Section \ref{subsec:active_passiveYM}  I will perform the same  analysis for Yang-Mills theory.  Lastly, in Section \ref{sec:gloss}, I will conclude that this passive-active correspondence implies that the local dynamical symmetries of both Yang-Mills and general relativity can be construed as notational redundancies.

\subsection{The passive-active correspondence for spacetime diffeomorphisms}\label{subsec:passive_active_GR}

I here define charts as smooth maps from subsets $U$ of $M$ (whose union covers $M$), to $\RR^n$, with smooth inverses.  The charts are are required  to have smooth transition functions wherever they overlap: given $\phi_1, \phi_2:U\rightarrow \RR^n$, where $U$ is the intersection of the domains of $\phi_1, \phi_2$, we require that $\phi_2\circ \phi_1^{-1}$ is a smooth bijective function between  subsets of $\RR^n$, from $\phi_1(U)$ of $\RR^n$ to $\phi_2(U)$. 
Any such complete collection of charts  is called \emph{an atlas} for $M$, and any two compatible atlases---whose transition functions between charts of the two atlases are smooth and have smooth inverses---are equivalent. The smooth structure of the manifold is defined as the equivalence class of atlases; or equivalently, as the maximal atlas, including all compatible charts. A maximal atlas can be taken simply to  \emph{define} the smooth and topological structure of the manifold.

First, let us look at the active transformations, as they act on charts. For $f\in \Diff(M)$ and a given tensor field $\mathbf{T}:=T^{a_1, \cdots, a_k}_{b_1, \cdots, b_l}$, we obtain a transformed field $\tilde{\mathbf{T}}:=f^*\mathbf{T}$: the `dragged' version of the tensor field. Of course, any chart that is dragged by a diffeomorphism also gives another chart. So, suppose that, under a chart $\phi_1:U_1\rightarrow \RR^n$, the components of $\mathbf{T}$ at a point that lies in $\phi_1$'s domain are given by $T^{\mu_1, \cdots,\mu_k}_{\nu_1,\cdots,  \nu_l}$. Then, there will be a second, compatible  chart, $\phi_2:U_2\rightarrow \RR^n$, for which the components of  the transformed field, $\tilde{\mathbf{T}}$, are \emph{also numerically given} by the \emph{untilded}  $T^{\mu_1, \cdots, \mu_k}_{\nu_1, \cdots, \nu_l}$. The relation between $\phi_1$ and $\phi_2$ is, of course, just  $\phi_1=\phi_2\circ f$, where $U_2=f(U_1)$. That is: 
\be
\begin{tikzcd}
    U_1 \arrow{r}{f} \arrow[swap]{dr}{\phi_1} & U_2 \arrow{d}{\phi_2} \\
     & \phi_1(U_1)\cap\phi_2(U_2)
  \end{tikzcd}
\ee 
Thus, given the joint description of $\mathbf{T}$ by all the charts $\{\phi_1^i\}_{i\in I}$ of  our atlas for $M$; call it atlas 1:  there will be a second atlas---atlas 2: $\{\phi_2^i\}_{i\in I}$---for which the \emph{different} tensor, $\tilde{\mathbf{T}}=f(\mathbf{T})$, has the same numerical description as $\mathbf{T}$. In equations: 
\be\label{eq:act_pass} \phi^i_1(\mathbf{T})=\phi^i_2(\tilde{\mathbf{T}}),\quad \forall i\in I.\ee
In  words, the images (i.e. the values of components) of the transformed tensor under the new charts  are the same as the images of the untransformed tensor under the old charts.\footnote{Here we opted for the standard construction of the manifold structure using charts to `probe' $M$. But there is a different route, that takes only the differentiable structure of $\RR^n$ for granted, and induces the smooth structure on $M$ from the bottom up. The interpretation is `nominalist' in the sense  that the charts are not understood as surveying some pre-existing abstract structure: they \emph{induce} the structure. A maximal atlas can be taken simply to  \emph{define} the smooth and topological structure of the manifold. In particular, one does not need to remain faithful to some prior   topological or smooth structure of $M$: the topology, as well as the
differentiable structure, are bequeathed to $M$  by the
charts of a maximal atlas.  The set of all domains of
charts in the atlas forms a topological base for the manifold: it is closed under finite intersections and arbitrary unions, and its union is the
whole manifold. With respect to this topology   all charts are  homeomorphisms, by construction. Cf. \citep[p. 22-23]{Lang_book} for a textbook definition of smooth structure in this manner,  and \cite{Wallace_coords} for a conceptual treatment.  Using the chart-nominalist interpretation of smooth structure,  the  fact that the domains of these charts in the two different atlases differ seems inconsequential, since the manifold structure (topological, smooth, etc) is \emph{defined} by the charts. \label{ftnt:chart_nom}  }

The active transformation therefore amounts to a change of (a non-maximal) atlas. And any diffeomorphism will leave a maximal atlas completely invariant; a good thing, since otherwise we would not be able to say that smooth structure---the structure that remains invariant under diffeomorphisms--- is determined (or defined, cf. footnote \ref{ftnt:chart_nom}) by a maximal atlas.

Now, a passive diffeomorphism is  a smooth function from  $\RR^n$ to itself, with smooth inverse, which I will write as $\bar f\in \Diff(\RR^n)$, that is interpreted as `translating' between two charts $\phi_1$ and $\phi_2$ of an atlas; so that  $\phi_2=\bar f\circ \phi_1$. Here we construe $\bar f$ passively, as a pure notational change:  when the domains of two arbitrary charts $\phi_1, \phi_2$  overlap, we have a transition function between the charts that is a diffeomorphism between subsets of $\RR^n$. That is: 
\be
\begin{tikzcd}
    \arrow{d}{\phi_1} U_1\cap U_2 \arrow{dr}{\phi_2} \\
    \phi_1(U_1\cap U_2)  \arrow{r}{\bar f} & \phi_2(U_1\cap U_2) 
  \end{tikzcd}
\ee 
From the above,  
\be \bar f:=\phi_2\circ \phi_1^{-1}:\phi_1(U_1\cap U_2)\subset\RR^n\rightarrow \phi_2(U_1\cap U_2)\subset \RR^n.\ee  This transformation simply does not act on quantities on $M$: we interpret it as only changing their description.\footnote{ When the domains overlap, one could see an active diffeomorphism as a right action of the diffeomorphisms on the charts, $\phi_2=\phi_1\circ f$, whereas the passive diffeomorphism above would correspond to a left action of the diffeomorphisms on the charts  $\phi_2=\bar f\circ \phi_1$.} 

Clearly, in order to reconceive $\bar f$ actively,  we can uniquely `associate' it  to \emph{a  local} diffeomorphism  on $M$, the domain manifold. More explicitly: given the charts,  we can reconstruct an active diffeomorphism relating a tensor ${\mathbf{T}}$ to a tensor $\tilde{\mathbf{T}}$ on a patch, using the transition function $\bar f$.  Namely,  by going down to $\RR^n$ by the chart, applying $\bar f$, and then going up from $\RR^n$ by the same chart. That is, omitting for now the index $i$:
\be\label{eq:pull_atlas} f(\mathbf{T})=: \tilde{\mathbf{T}} \quad\text{where}\quad f:=\phi_1^{-1}\circ \bar f\circ\phi_1\in \Diff(M).\ee 

Most authors would be wary of identifying active and passive transformations in such an explicit fashion: for one thing, they will point out, active diffeomorphisms act globally, whereas passive transformations act on each chart. 

But first of all,  it is undeniable that tensor quantities $Q$ such that $\phi_1^*(Q)$  are invariant under the passive $\bar f$. That is, using the fact that in our notation the pull-back works as the push-forward (cf. footnote \ref{ftnt:subs_sups}) and that:
\be\label{eq:pass_inv}\bar f^*(\phi_1^*(Q))=\phi^*_1(Q),\ee this tensorial quantity will also be  invariant under the active  diffeomorphism $f$ given in \eqref{eq:pull_atlas}, i.e.:
\be f^*(Q)=\phi_1^{-1}{}^*\circ \bar f^*\circ\phi_1^*(Q)=\phi_1^{-1}{}^*\phi_1^*(Q)=Q,\ee
since \eqref{eq:pass_inv} also implies that $\bar f^{-1}(\phi_2(Q))=\phi_2(Q)$. 

Conversely, any active \emph{infinitesimal} diffeomorphism, represented by the vector field $X^a$, will, in the infinitesimal limit, map points not in the boundary of each chart to points within that same chart, and assuming charts overlap on open subsets, the boundary of each chart will belong to the interior of another chart, etc. Then, by the same construction that gave rise to \eqref{eq:pull_atlas}, these active infinitesimal diffeomorphisms  correspond to some infinitesimal diffeomorphism of each chart in the given atlas, transitioning in the appropriate way at the intersection of neighboring charts. 

Therefore, there is a 1-1 correspondence between  quantities that are invariant under the diffeomorphisms that are connected to the identity---that is, that are generated by the flows of vector fields---and those quantities that are invariant under   coordinate transformations that are connected to the identity.  
Intuitively, this relation is mathematically rather simple, since we know that local patches of $M$ are locally diffeomorphic  to $\RR^n$, and we can therefore naturally move diffeomorphisms from one space to the other.\footnote{ For illustration, take the change from spherical to cylindrical coordinates: $$ r=\sqrt{\rho^2+z^2}; \,\, \theta=\arctan\left(\tfrac{z}{\rho}\right); \,\, \varphi=\varphi, $$
where $r$ and $\rho$ are, respectively, cylindrical and spherical radius, $\varphi$ is the azimuth angle, $z$ is the cylindrical  height, and $\theta$ is the elevation angle. Passively, we take these coordinates to refer to the same points of $\RR^3$. But we can also construe this diffeomorphism actively, for $\RR^3$ comes endowed with some background structure of its own. So the map above takes a given ordered triple, seen as an element of the product $\RR^3$,  to a different ordered triple: just plug in values of $(\rho,  \varphi, z)$ and find where they go as $(r(\rho,z), \varphi, \theta(z,\rho))$. Under this interpretation, a given $(a,b,c)\in \RR^3$ is being \emph{actively} mapped to $(\sqrt{a^2+c^2}, b, \arctan\left(\tfrac{c}{a}\right))\in \RR^3$.  }   But despite this simplicity, to the extent that we think passive transformations are better understood or at least more operational (as I will argue in \cite[Sec. 5]{Samediff_1b}), this relation provides a powerful interprettive tool.

In sum: there are two ways of associating active and passive transformations. The first way says that an active diffeomorphism takes quantities as described by one   atlas to the same description  under a different, compatible atlas, as in equation \eqref{eq:act_pass}.   The second association between active and passive is more useful, since it does not require us to explicitly change the atlas by reshuffling. For instance, for diffeomorphisms,  it says that   passive diffeomorphisms---diffeomorphisms of $\RR^n$ on the image of the charts---recover some of the active diffeomorphisms (cf. \eqref{eq:pull_atlas}).  In particular, we get a 1-1 correspondence between  the active and passive infinitesimal symmetries of general relativity (as defined in Section \ref{sec:syms}), and therefore, by integrating in time, we get a 1-1 correspondence between `invariants under coordinate transformations that are connected to the identity' and `invariants under isomorphisms that are connected to the identity'.\footnote{Thus if we want to construe symmetry just as notational variance, we have strong reason to restrict considerations to  symmetry groups that are connected to the identity---as they appear in the Hamiltonian formalism, as I argued in Section \ref{sec:syms} (cf. also \cite[Sec. 3]{GomesButterfield_electro}). But I should note that the mismatch between the full group of diffeomorphisms and the subgroup that is connected to the identity is stark: according to one suitable topology---called `weak Whitney topology'---any open set of $\Diff(M)$  containing the identity also contains elements that are not connected to the identity. Another reason to restrict to the symmetry groups that are connected to the identity, as we will discuss in \cite{Samediff_2}, is that in the Hamiltonian framework, symmetries are generated by the symplectic flow of constraint functions, or momentum maps, and are thus always connected to the identity. Indeed, I take these latter facts to justify the physicist's focus  on invariance with respect to coordinate transformations, as opposed to the more abstract invariance under active diffeomorphisms. \label{ftnt:con_id}}

\subsection{Active and passive correspondence for gauge transformations}\label{subsec:active_passiveYM} 
As with the definition of a manifold using an atlas,   the intrinsic construction of bundles in Section \ref{sec:PFB}       ``hides under the hood" the fact that we can {\em define} bundle structure using ideas about local trivializations. Namely,  we use local trivializations  and  conditions on the transition functions between charts to induce the  bundle structure from a local product structure. And once again, the invariant structure is defined by what are taken to be `coordinate tranformations'.

A local section of a principal bundle $P$, $\sigma$, induces a diffeomorphism $U\times G\simeq \pi^{-1}(U)$, given by $\Sigma:U\times G\rightarrow P$, such that:
\be\Sigma: (x, g)\mapsto g\cdot {\sigma}(x), \quad\text{whose inverse is}\quad \Sigma^{-1}: p\mapsto (\pi(p), g_{\sigma}(p)^{-1})\ee 
where $g_\sigma:\pi^{-1}(U)\rightarrow G$ gives   $g_{\sigma}(p)$ as the unique group element taking $p$ to (the fibre's intersection with) the local section, i.e. $g_\sigma(p)$ is the group element such that 
\be\label{eq:sbar}g_{\sigma}(p)\cdot p={\sigma}(\pi(p)).\ee
The precise form of $g_{\sigma}$ will of course depend on ${\sigma}$.

Vertical automorphisms $\tau$, given in \eqref{eq:tau}, can be represented  with a group-valued function on $P$, i.e. $\Psi:P\rightarrow G$,  defined by 
\be\label{eq:Psi}\tau(p)=\Psi(p)\cdot p\quad \text{such that}\quad  \Psi(g\cdot p)=g\Psi(p)g^{-1},\ee which is the  equivariance condition that $\Psi$ gets from $\tau$.
  
 Then any vertical  automorphism $\tau$ induces a diffeomorphism of $U\times G$, as follows. Let $\tau(p):=\Psi(p)\cdot p$, as above.  Then, for a section ${\sigma}$ and  a general $p=\Sigma (x,g)\in  \pi^{-1}(U)$,  
using \eqref{eq:sbar} gives: 
 \be  \tau\circ \Sigma: (x,g)\mapsto \tau(g\cdot \sigma(x))=\Psi(g\cdot \sigma(x))\cdot(g\cdot {\sigma}(x))=(\Psi(g\cdot \sigma(x))g)\cdot {\sigma}(x),
 \label{eq:taup}\ee
where in the first equality we used the definition of $\Psi$ in \eqref{eq:Psi} and in the second equality we used that the group action is a homomorphism: $h\cdot (g\cdot p)=(hg)\cdot p$, where $h,g\in \G$ (and above, $h=\Psi(g\cdot \sigma(x))$ and $p=\sigma(x)$).

 As expected, the vertical automorphism $\Psi$ just takes $\sigma$ to a different section, which, taking $g=\mathrm{Id}$ in \eqref{eq:taup}, is immediately seen to be $\sigma':=\Psi(\sigma)\cdot \sigma$.
 Moreover, applying $\Sigma^{-1}$ to \eqref{eq:taup}, we obtain, since $\Sigma^{-1}(g\cdot \sigma(x))=(x, g)$:
  \be\label{eq:auto_gt}\Sigma^{-1}\circ\tau\circ \Sigma: (x, g)\mapsto (x, \Psi(g\cdot \sigma(x))g)=(x, g\Psi({\sigma}(x))),\ee
  where we used the equivariance property of $\Psi$ of \eqref{eq:Psi} on the last equality. And so, sandwiched between the diffeomorphism $\Sigma$, the vertical automorphism only acts on the group part of the product $U\times G$, with $g^\sigma_\Psi:=\Psi\circ {\sigma}:U\rightarrow G$. We thus obtain that $\Sigma^{-1}\circ\tau\circ \Sigma$ is a `coordinate transformation', or diffeomorphism of $U\times G$.\footnote{ Since  ${\sigma}$ and $\Psi$ are smooth, and, for fixed $x$,  $g\mapsto g\Psi({\sigma}(x))$ is clearly a diffeomorphism of $G$ (since it is just the action of $G$ on the element $\Psi({\sigma}(x))$). The inverse is of course just $(x,g)\mapsto (x, g\Psi({\sigma}(x))^{-1})$, which enjoys the same properties.}
     
 We call a $g^\sigma_\Psi\in \G$ a \emph{gauge transformation}. To be defined,   these gauge transformations require a trivialization, $\sigma$. Once anchored to a trivialization, they  are the local, passive counterparts of the active $\Psi:P\rightarrow G$, described in \eqref{eq:Psi}. The set of $g^\sigma_\Psi$ for all vertical automorphisms $\Psi$  defines $\G:=\{g(x), \,\,x\in U\}$, which inherits from $G$ the structure of an (infinite-dimensional) Lie-group, by pointwise extension of the group multiplication of $G$ over $U$.

Can we understand these passive transformations, these changes of sections $\sigma$, in terms of more familiar mathematical objects, as we understood the coordinate changes of the spacetime manifold? Yes, we can understand them as point-dependent changes of bases for vector spaces, in two ways. One is more indirect, which I include here in this Section;  and the other is more direct, since it is based on a spacetime representation of the Ehresmann connection, to be described in Appendix \ref{sec:Atiyah}. 

As discussed in Section \ref{subsec:PFB_ex}, given some general vector space $F$ and structure group $G$ and $\rho:G\rightarrow GL(F)$, and $P$ a $G$-principal bundle over $M$,  we can define the associated vector bundle  over $M$, which is denoted $E:=P\times_\rho F$. Conversely, the frame bundle for a given vector bundle $E$, $L(E)$ (formed by the bases of $E_x\simeq F$ for each $x\in M$) is a principal bundle $P'$ with structure group  $GL(F)$. But we can form another principal bundle $P$, as a sub-bundle of $P'$ as a sub-bundle of $L(E)$ corresponding  to a subset of frames   related by $\rho(G)$.  Then, assuming the action of $\rho:G\rightarrow GL(F)$ is faithful, changes of section in $P$ uniquely  correspond to changes of frames of $E$; cf. \citet[Prop. 5.5]{kobayashivol1} and \citet[p. 2401]{Weatherall2016_YMGR} for a conceptual treatment.\footnote{Of course, this raises a puzzle: if the principal bundle is construed as just a bundle of linear frames, how can we justify  the restriction of $G$ to a subset of the most general group of transformations between frames, $GL(F)$? As discussed by \cite[Sec. 4]{Weatherall2016_YMGR},  the restriction corresponds to the preservation of some added structure to $F$. In other words, when $F$ is not just a  vector space, but e.g. a normed vector space, we would like changes of basis to preserve this structure, e.g. the orthonormality of the basis vectors, and this restricts the bundle of linear frames to the appropriate sub-bundle. To see this, define $P\times_\rho F$ as the equivalence class for the doublet $(p, v)\in P\times F$  with  $(p, v)\sim (g\cdot p, \rho(g^{-1})v)$. Suppose  that $F$ is a Riemannian vector space, with metric $\langle\cdot,\cdot\rangle$. We can  induce a metric in $P_F=P\times_GF$  defining, for any  $p$ and $v,v'\in F$:
$\langle[p,v],[p,v']\rangle:=\langle v,v'\rangle$.
To be well-defined, we must have:
$$\langle[p,v],[p,v']\rangle=\langle[g\cdot p,\rho(g^{-1})v],[g\cdot p,\rho(g^{-1})v]\rangle=\langle \rho(g^{-1})v,\rho(g^{-1})v'\rangle:$$
which is true only if the action of the group on $F$ is orthogonal with respect to the metric. This corresponds to $G=O(n)$; similarly, $SO(n)$ adds an orientation to $F$. Similarly, $G=U(n)$ corresponds to a complex vector space structure and a Hermitean inner product; $G=SU(n)$ adds an orientation (see \cite[p. 60]{kobayashivol1} and \cite[p. 2403]{Weatherall2016_YMGR} for a conceptual treatment). The moral is that the added structure on $F$ induces an added structure on  the associated vector bundle only if the  transformation group preserves that added structure. 
}


In other words, a section $\sigma:U\rightarrow P$ may be understood as a frame field for a certain vector bundle, and changes of section may be understood as the allowed change of basis at each point.

\subsection{Notational redundancy}\label{sec:gloss}

  Thus, by having the symmetries (according to Definition \ref{def:inf_sym} of Section \ref{sec:syms}) of our theories---general relativity and Yang-Mills---be induced by the automorphisms of a natural geometric structure---smooth manifolds and smooth fibered manifolds, respectively---we have achieved something remarkable: both of these symmetries can now be glossed as mere notational redundancy.


This construal closes a gap between the more conceptual-minded physicist or philosopher of physics and the more pragmatic physicist. For the latter,  invariance under different coordinate representations is usually \emph{equated} with `physical status'. As \citet[p. vii]{Dirac_QM} writes in the introduction to his magisterial book,  : 
``The important things in the world appear as the invariants
[...] of these [coordinate] transformations.''  And here is 
\citet[p. 82]{Nozick}, espousing a complete disregard for active transformations: ``Once we possess the covariant representation under which the
equations \emph{stay the same for all coordinate systems}, the quantities in the (covariant) equations are the real and objective quantities.''  (my emphasis).\footnote{Note here that, though Nozick focuses on coordinate transformations, he is not requiring the physical quantities to remain invariant under these transformations. So tensorial quantities would, for Nozick, qualify as real.   }

In both the general relativistic and Yang-Mills case, this property---that the active isomorphisms are locally equivalent to a passive transformation---gives a gloss of `notational redundancy' to the  symmetry in question, 
 a type of redundancy  most authors agree to be well understood.

Thus \citet[p. 2404]{Weatherall2016_YMGR} writes:  
 \begin{quote}
 We are thus led
to a picture on which we represent matter by sections of certain vector bundles (with additional structure), and the principal bundles of Yang–Mills theory represent various
possible bases for those vector bundles. 
 These considerations lead to a deflationary attitude towards
notions related to ``gauge'': a choice of gauge is just a choice of frame field relative
to which some geometrically invariant objects [...] may be represented, analogously to how geometrical objects may be
represented in local coordinates.\end{quote}
As I have argued, invariance under coordinate change can only play this deflationary role once an active-passive correspondence for the symmetries of the theory is established, as it was here.

In \cite{Samediff_1a}, one desideratum for the symmetries of any given physical theory to admit a `sophisticated' interpretation, in the same sense as---I will argue---general relativity and Yang-Mills theory do, will be precisely that the theory's symmetries be induced by the automorphisms of a natural geometric structure. For once symmetries are construed as isomorphisms that are  induced by the  automorphisms of a natural geometric structure, and we can build an active-passive correspondence as above, we can---as Weatherall says above---understand symmetry-invariance as mere notational invariance under coordinate changes. Thus, for theories that satisfy that desideratum, we are motivated to construct a passive-active correspondence, which would indeed reinforce the common belief that the symmetries of those theories correspond to notational redundancy.



\section{Summing up }\label{sec:conclusions_soph}

 Yang-Mills and general relativity are  the two theories that underlie our most empirically successful models of the world.  Among the most philosophically contentious topics about these theories is the relationship between symmetry and physical equivalence. With the current  series of papers, \cite{Samediff_1a, Samediff_1b, Samediff_2}, of which this is the first, I hope to shed light on that topic, from various different angles. In the process, we will clarify how the symmetry-invariant structure of each theory is to be mathematically and physically understood. 
 
 To start my investigation,  here I provided  specific definitions of symmetries on state space. The definition of infinitesimal  symmetry was then  applied to whatever mathematical object was responsible for endowing the theory with dynamics: e.g. the action functional or the Hamiltonian. These we called \emph{dynamical symmetries}, which,  following \cite{Wallace2019},  we argued to be empirically unobservable.    
 
 In most theories, one lists a set of dynamical symmetries that fit under the given definition. But seldom is an exhaustive list provided, leaving open the possibility that ``unwanted'', un-listed, symmetries persist in the theory.  Here, for both the Yang-Mills and the general relativistic case, I provided an exhaustive list of dynamical symmetries under the present definition. 
 
 Unlike the dynamical symmetries of general relativity, which are induced by the automorphisms of the underlying spacetime manifold,  at first sight the dynamical symmetries of Yang-Mills have no geometric significance; they are postulated. But  I showed how these symmetries can also be construed as induced by the automorphisms of a  natural geometric structure:  as fiber-preserving diffeomorphisms of an appropriate fibered manifold. 

Thus we showed that the symmetries of ${A^I_\mu}$ are in 1-1 correspondence with the symmetries of $\omega$, which have a geometric origin. I then proved the same correspondence holds for gravity:  namely, the active isomorphisms of $g_{ab}$ that are connected to the identity are in 1-1 correspondence with the passive coordinate changes---also connected to the identity---of all the local coordinate representatives $g_{\alpha\beta}$. Thus the invariants functions of of $\omega$ and $g_{ab}$ under  infinitesimal dynamical symmetries  can be understood purely passively: as invariants under coordinate changes. The conclusion is that, when the isomorphisms of the theory are induced by the automorphisms of an underlying geometric structure---a smooth manifold or a fibered smooth manifold---we can ``deflate'' the meaning of symmetries to mere notational changes; like the change of basis of a vector space. 

 But the relationship between $g_{ab}$ and $g_{\alpha\beta}$ is not the same as the relationship between $\omega$ and ${A^I_\mu}$. In the metric case, both the abstract tensor and the coordinate expression have spacetime as their domain. The relation between $\omega$ and ${A^I_\mu}$ requires a pull-back to a section, which is why we can interpret the latter but not the former as a function over spacetime. The closer parallel therefore is between a section $\Gamma$ of the Atiyah-Lie algebroid, described in Section \ref{sec:Atiyah}, and ${A^I_\mu}$. For the sections $\Gamma$ are in 1-1 correspondence with the Ehresmann connections, $\omega$, but they are abstract sections of a vector bundle over spacetime, which are represented as ${A^I_\mu}$ once we choose a coordinate basis for spacetime and for each copy of the  Lie algebra over each point of spacetime in a chart.

\subsection*{Acknowledgements}

I would like to thank Jeremy Butterfield for many conversations on this topic, and for reading and commenting on various versions of the paper. I would also like to thank Ruward Mulder, for a careful reading and insightful questions. And I would also like to thank two anonymous referees, for their patience and   comments, which led to several revisions of the initial draft.

\appendix

\section*{APPENDIX}

   \section{The bundle of connections}\label{sec:Atiyah}

Note above that the basic field that lends itself to the geometric interepretation, namely, the Ehresmann connection, $\omega$, is \emph{not} a field on spacetime, but on some other (fibered) manifold. And the interpretation of passive transformations as changes of frames above required some vector bundle, $E$ (or indeed, mostly the model vector space, $F$), representing the value space of some other field that interacted with the force carried by $\omega$. Can we interpret the passive transformations as of an interacting field in spacetime,  without tying it to a specific matter field with which it interacts?  To finish the side-by-side comparison of Yang-Mills and general relativity, we would like to describe the  Yang-Mills fields as on a par with the abstract metric tensor field, without the use of coordinates and as fields on spacetime, and to understand the passive gauge transformations as changes of bases for the values of this field.

Here I introduce the bundle of connections and their sections, also known as connections of the Atiyah-Lie bundle, as a global, spacetime representative of the connection-form.\footnote{The bundle of connections appeared almost simultaneously in  \cite{AtiyahLie} and  \cite{Kobayaschi_bundle}. It is often  referred to as the \emph{Atiyah-Lie bundle}. See also \cite[Ch. 17.4]{Kolar_book}. To avoid confusion, it is better to refer to a section of the bundle of connections, which is itself a generalization  of a connection to what are known as Lie algebroids (see \cite{mackenzie_2005}), as an Atiyah-Lie connection.}

To start, we could ask what kind of bundle the gauge potentials  $\mathbf{A}$ are sections of. This would lead us to a global section of some vector bundle over spacetime, which would reduce to $\mathbf{A}$ in particular coordinate patches. But before we do that, note that $\mathbf{A}$ is  object that mixes tensorial indices with internal indices. The natural principal bundle for the tensorial part, as discussed above (see \cite[Sec. 3]{Weatherall2016_YMGR}), would be a sub-bundle of the  frame bundle $L(TM)$. The internal part, corresponding to $\mathfrak{g}$, would require a sub-bundle of $L(P\times_\rho\mathfrak{g})$.  Here $\rho=\mathrm{Ad}:G\rightarrow GL(\mathfrak{g})$,   where $\mathrm{Ad}_g v=g^{-1}v g$ is the natural, adjoint action of $G$ on $\mathfrak{g}$, appearing in \eqref{eq:omega_defs} and \eqref{eq:gauge_trans}). 

To define the global sections of such an object, one would need to \emph{splice} bundles of these different characters together (see e.g. \cite[Ch. 7.1]{Bleecker}). Although it is possible to construct the bundle in this way, it would involve the introduction of yet more formalism.  But there is an alternative way, that leads to the same answer, and which I will now explain (see the Proposition in \cite[Ch. 17.5]{Kolar_book}, for their equivalence). 



 Parallel transport is determined by horizontal directions in the bundle, as we saw in Section  \ref{sec:PFB}, and  we know that the horizontal bundle $H\subset TP$, is left-invariant.  So, if we know what parallel transport is at $p$, we know what it is at $g\cdot p$. By getting rid of this redundancy, we can find a global spacetime representation of the connection $\omega$. 
To do that, we first note that there is a 1-1 relation between (Ehresmann) connection-forms and \textit{left-invariant} sections of $TP$ (see \cite[Ch. 4]{kobayashivol1}). 

Left-invariant vector fields are not unconstrained sections of the vector bundle $TP$, i.e. $C^\infty(TP)$. But they \emph{are} unconstrained sections  of $TP/G$, the so-called \textit{bundle of connections} (see e.g. \cite[Sec. 3.2]{Ciambelli}; \cite[p.9]{LeonZajac}; \cite[p.60]{sardanashvily2009fibre}; \cite[Ch. 17.4]{Kolar_book} and \citep[Ch. 7]{Jacobs_thesis}). In other words, the difference between sections of $TP$ and $TP/G$  is that, while both can be seen as sections over $TM$ (with $\pi_*$ the projection), the latter---$TP/G$---is more constrained, since it can only encode left-equivariant objects defined on the first, $TP$.

The main idea in the construction of this bundle  is to take the projection map $\pi_*:TP\rightarrow TM$, and make it `forget' at which point or ``height'' of the orbit   it was applied.  The formalism  represents parallel transport of internal quantities for the directions in spacetime, rather than for directions in the bundle $P$.  
Thus $TP/G$ is most naturally a vector bundle over $TM$ rather than over $M$ or  $P$. But since $TM$ is itself a bundle over $M$, $TP/G$ can  also be construed as a bundle over $M$.

To define the fiber of $TP/G$, recall that a point in $TP$ is
locally described by $(p, v_p)$ with $v_p \in T_pP$, and the group $G$ acts (freely and transitively)  as
$(p, v_p)\mapsto (g\cdot p, {L_g}_*(v_p))$, which is the relation by which we define the left-invariant vector fields. Thus $TP/G$ is defined by identifying
\be(p, v_p) \sim (g\cdot p, {L_g}_*(v_p)), \quad \text{for all}\quad g\in G.\ee
Since locally (i.e. given some trivialization of the tangent  bundle) for $x=\pi(p)$ and $\xi\in \mathfrak{g}$, we can represent $p=(x, g):=g\cdot \sigma(x)$ and $v_p=(X_{x}, \xi):=\xi+\sigma_*(X_x)$ we have, locally, $(p, v_p)=(x,g, X_x, \xi)$. 
If we take the quotient, we obtain that the elements of the new vector bundle will be locally of the form $(x, X_x, \xi)$.  


Given a point on $M$, and a tangent direction on $M$, and a local trivialization of the bundle, an element of the vector bundle $T^*P/G$  spits out a Lie-algebra element. Thus, as in the standard manner of obtaining $\mathbf{A}^\sigma$ from $\omega$, here we also locally recover, in a trivialization, that the representative of the connection, call it $\Gamma$,  is the $\mathfrak{g}$-valued 1-form on $M$; 
$\Gamma$ is global, but in a local trivialization, it would be represented by $A_\mu^I$, where, the indices refer to a Lie-algebra and a tangent bundle basis.  So $\Gamma$ stands to the abstract tensor $g_{ab}$ as $A_\mu^I$ stands to  $g_{\mu\nu}$.
 The values of $\Gamma$ according to different trivializations are  related by the transformation \eqref{eq:gauge_trans}, just as the values of $g_{\mu\nu}$ are related by coordinate transformations. These are correlates of  the passive transformations.  Thus we find, as announced in the introduction to this Section, the appropriate analogy, comparing a section of $TP/G$ with a global vector field, $\mathbf{X}$, which we can write locally with coordinates, $X^\mu\pp_\mu$, where $\mathbf{A}^\sigma$ stands in analogy to the components $X^\mu$. Thus, the sections of the bundle $TP/G$ will be frame-invariant, and therefore, invariant under passive gauge transformations.

 We can sum up  as follows: a section of $T^*P/G$ should be seen as the global, coordinate-independent generalization of $\mathbf{A}^\sigma$; the advantage of a section of $T^*P/G$ over  the standard gauge potential is that it is globally defined and it is independent of internal coordinates (coordinates for the Lie algebra, and tangent bundle); and the advantage over the connection-form is that it is a section of a vector bundle with $T^*M$ as its base space. The disadvantage is that it is highly abstract. Nonetheless, this formulation allows a strong analogy between the basic kinematical variables of the gauge theory and the metric, in a coordinate-independent manner.

To finish  this Appendix, let us briefly focus once  again on the geometrical meaning of $\omega$. The unifying power of the principal connection is that it defines compatible parallel transport for any field/particle that interacts with the force associated to $G$, even for the as-of-yet undiscovered forces and groups. 

We can think of $\Gamma$, the section of the vector bundle $T^*P/G$, as one more physical field on spacetime. Since it is a section of a certain vector bundle, upon introducing coordinates (or frames) it admits changes of bases with which it is described, and these can be construed as passive gauge transformations. Just as the connection $\omega$ is invariant with respect to these passive transformations, but variant with respect to the active ones, so will be $\Gamma$.

 \end{document}